\documentclass[final,5p,times,twocolumn]{elsarticle}

\usepackage{amssymb}
\usepackage{float}
\usepackage{hyperref}
\usepackage{color}
\usepackage{amsmath}
\usepackage{lineno}

\journal{Nuclear Instruments and Methods in Physics Research A }

\begin{document}

\begin{frontmatter}



\title{HIBISCUS: a new ion beam radio-frequency quadrupole cooler-buncher for high-precision experiments with exotic radioactive ions}


\author[inst1,inst2]{A. Jaries}

\affiliation[inst1]{organization={University of Jyvaskyla, Department of Physics},
            addressline={Accelerator Laboratory, PO Box 35}, 
            city={FI-40014 University of Jyvaskyla},
            country={Finland}}

\affiliation[inst2]{organization={Helsinki Institute of Physics}, addressline={FI-00014, Helsinki}, country={Finland}}

\author[inst1]{J. Ruotsalainen}
\author[inst1]{R. Kronholm}
\author[inst1]{T. Eronen}
\author[inst1]{A. Kankainen}

\begin{abstract}
HIBISCUS (Helium-Inflated Beam Improvement Setup that Cools and Undermines Spreads),
a new radiofrequency quadrupole cooler-buncher device has been developed and commissioned offline at the Ion Guide Isotope Separator On-Line (IGISOL) facility in Jyväskylä in Finland, as an in-kind contribution for the Facility for Antiproton and Ion Research facility. 
HIBISCUS improves the ion optical properties of incident low-energy 6~keV beams with the option to have it ultimately extracted in temporally short bunches ($<1$~$\mu$s). This paper provides technical descriptions of its main characteristics, along with a set of optimum working parameters and performance in terms of transmission efficiency, longitudinal energy spread of the cooled ions and temporal width of the extracted bunches.
\end{abstract}




\end{frontmatter}



\section{Introduction}
\label{sec:sample1}
\begin{figure*}[htb!!]
\includegraphics[width =1\textwidth]{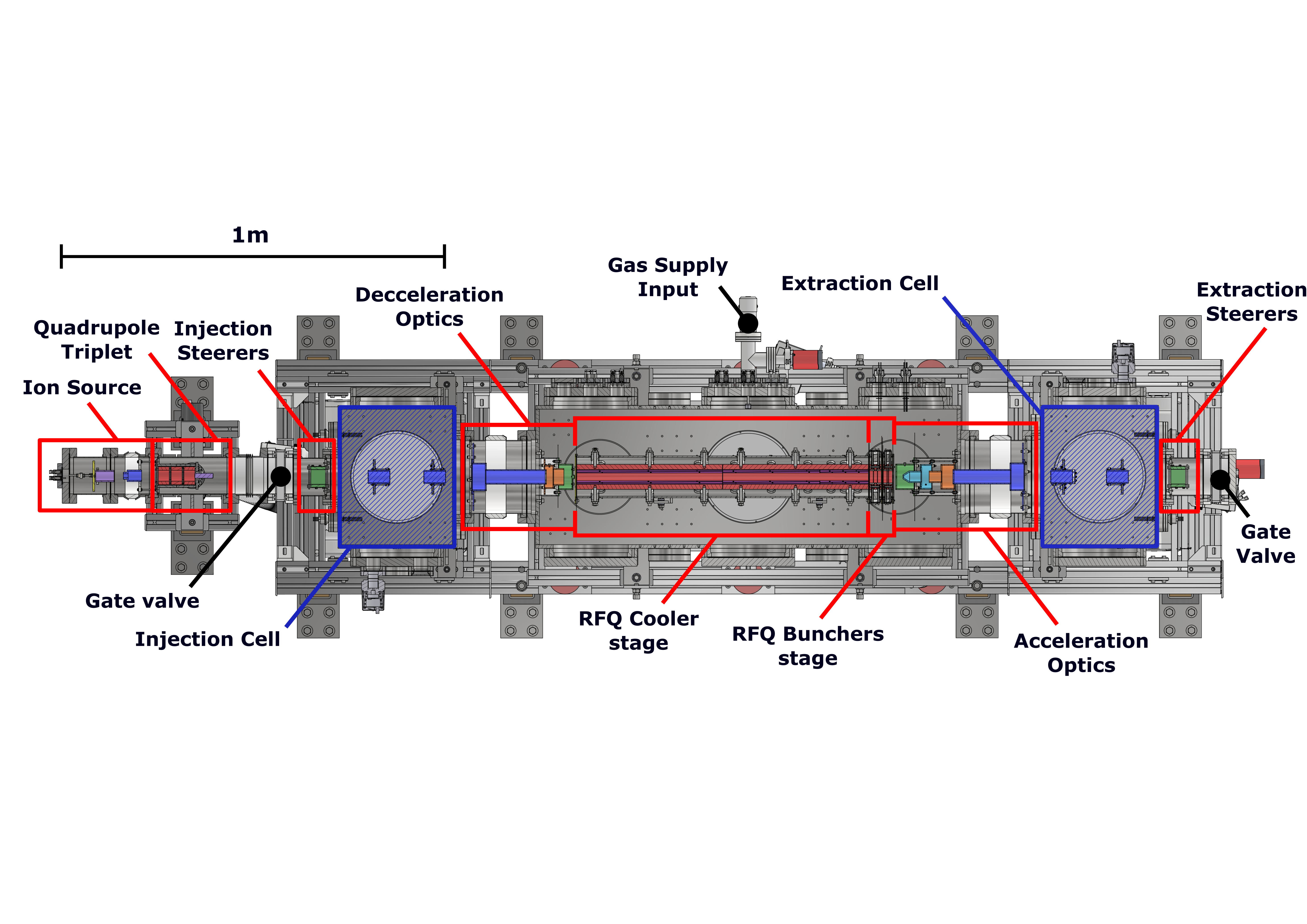}
\caption{Top overview of the HIBISCUS beamline. The different parts of the experimental setup are identified. The injection and extraction cells, marked with hatched blue boxes, are designed to house beam diagnostics. The sub-assemblies, discussed in more details in Sec.\ref{exp}, are marked with red boxes.\label{fig:beamline}}
\end{figure*}

As the field of experimental nuclear physics develops, ion beam radiofrequency quadrupolar cooler-bunchers (RFQCB) have become integral parts in modern radioactive ion beam (RIB) facilities, such as in IGISOL \cite{nieminen2001beam,nieminen2002online,eronen2014jyfltrap}, ISOLDE at CERN \cite{mukherjee2008isoltrap,mane2009ion,catherall2017isolde}, SHIPTRAP at GSI \cite{block2005ion}, TITAN at TRIUMF \cite{brunner2012titan}, CPT at ANL \cite{savard1997canadian}, ISOL-RISP \cite{boussaid2017technical}, LEBIT at NSCL \cite{schwarz2016lebit} or in DESIR at GANIL \cite{GERBAUX2023167631}.

The upcoming Facility for Antiproton and Ion Research \cite{spiller2006fair} (FAIR) facility, in Darmstadt, Germany, and its Nuclear Structure, Astrophysics and Reactions \cite{nilsson2015nustar} (NuSTAR) pillar will offer completely new possibilities to study exotic nuclei and their reactions, crucial for understanding processes in stars and elemental nucleosynthesis in the Universe. The Super-conducting FRagment Separator \cite{winkler2008status} (Super-FRS) will provide intense beams for studying the structure and dynamics of very exotic nuclei, accessing much further out in beam availability and intensity than achievable in currently existing facilities. Its low-energy branch will deliver ion beams with keV-scale energies to various experimental setups, such as Penning trap mass spectrometry (see Ref.~\cite{Eronen2016,dilling2018penning} and references therein) or collinear laser spectroscopy (see Ref.~\cite{campbell2016laser} and references therein), for which an RFQCB is an essential device allowing to suitably shape the incoming ion beams.

The HIBISCUS has been designed to deliver a beam with suitable ion optical properties and temporal bunch length while retaining high transport efficiency. It is composed of two main stages: a main cooling stage followed by a bunching stage, enclosed and separated by plates with narrow apertures to hold the low-pressure helium gas inside. Gas is fed only to the main cooling stage while the pressure in the bunching stage is dictated by the flow through the aperture placed in-between. Unlike most of the existing RFQCB found in RIB facilities, HIBISCUS does not have longitudinally segmented electrodes in its RFQ cooler section. Instead, the radio-frequency (RF) and static (DC) components, used to radially confine and axially guide the ions, respectively, are decoupled and applied on a much smaller number of electrodes. A similar design is found in the LEBIT cooler buncher at NSCL \cite{schwarz2016lebit}, which uses pairs of wedge-shaped DC-only electrodes placed in between the RF quadrupolar rods. The main role of the second stage is to bunch the cooled ions. While the bunching can already be performed at the far end of the first stage, this second stage, designed with two subsequent RFQ buncher sections, is dedicated to further reduce the temporal width of the extracted ion sample at the time focus. 

A detailed description of the design of the HIBISCUS setup and its surrounding beamline, a summary of its modes of operation, a collection of relevant experimental studies that were performed and the main results obtained during the offline commissioning are reported hereafter.

\section{Experimental setup \label{exp}}
\subsection{General layout}
An overview of the experimental setup is shown in Fig.~\ref{fig:beamline}. Its footprint is around 2.6~m in length and 0.6~m in width and is mainly composed of three rectangular vacuum chambers. The device is located between two vacuum gate valves to allow vacuum isolation from the rest of the main beamline. The main vacuum chamber houses the quadrupole sections. The diagnostics cells, placed on the injection and extraction sides of the central region, house the detectors used for characterization and monitoring. In addition, beam shaping elements are housed in cylindrical chambers, namely a set of electrostatic steerers, mounted on each diagnostics cell, and a quadrupole triplet assembled on the injection side, prior to the gate valve. The main central vacuum chamber, its supporting aluminum frame, pumping system and gas line are at high-voltage (HV). It is enclosed with an HV protection cage for safe operation, and to isolate it from the ground, the structure is mounted on plastic HV insulators. Ceramic insulators are also used to separate it from the diagnostic cells and the gas supply that are at ground potential. 

Finally, for offline commissioning purposes only, an ion source is also incorporated to the front of the whole assembly (see Sec.~\ref{IS}). As the cylindrical vacuum chamber of the source is at HV, it is enclosed in its own safety protection cage and electrically isolated from the rest of the beamline with a ceramic insulator. The electric potential of the chamber defines the energy of the emitted ions. The chamber is lifted some tens of volts higher than the HV level of the main RFQCB section to allow ions to be injected into the RFQCB.

Following the ion source, there are hollowed earth-grounded cylinders on the beam axis inside the HV insulators to shield the ions' trajectory from any charging-up effects of the ceramic insulator. The electrical insulation between the electrodes is done either using Kapton foils, alumina or PEEK plastic. All in-vacuum electrical connections are made using single-core copper wire with Kapton coating graded for 10~kV insulation.

The electrodes composing the different sub-assemblies identified with red boxes in Fig.~\ref{fig:beamline} are held together with bolts and rods and form geometrically rigid entities imposed by small mechanical tolerances from the machining. Their relative alignment within the sub-assembly is hence given at a level of a few tens of $\mu$m. The sub-structures are then mounted and their placement adjusted inside their respective vacuum chambers. The quadrupole triplet and diagnostics chambers are leveled and aligned with respect to the central RFQCB chamber using spherical washers and adjustment rods. The diagnostic cells are mounted on rails for practicality. The whole assembly and the aluminum frames lie on supporting feet allowing  $\pm2$~cm vertical adjustment around the beamline default height, set at 1.2~m from the floor. The sub-structures of the apparatus are detailed in the following sections. The design of the RFQCB is based on ion transport simulations optimized using SIMION \cite{ruotsalainen2021design,dahl2000simion}.

\subsection{Ion source \label{IS}}
\begin{figure}[htb!!]
\begin{center}
    \includegraphics[width =1\columnwidth]{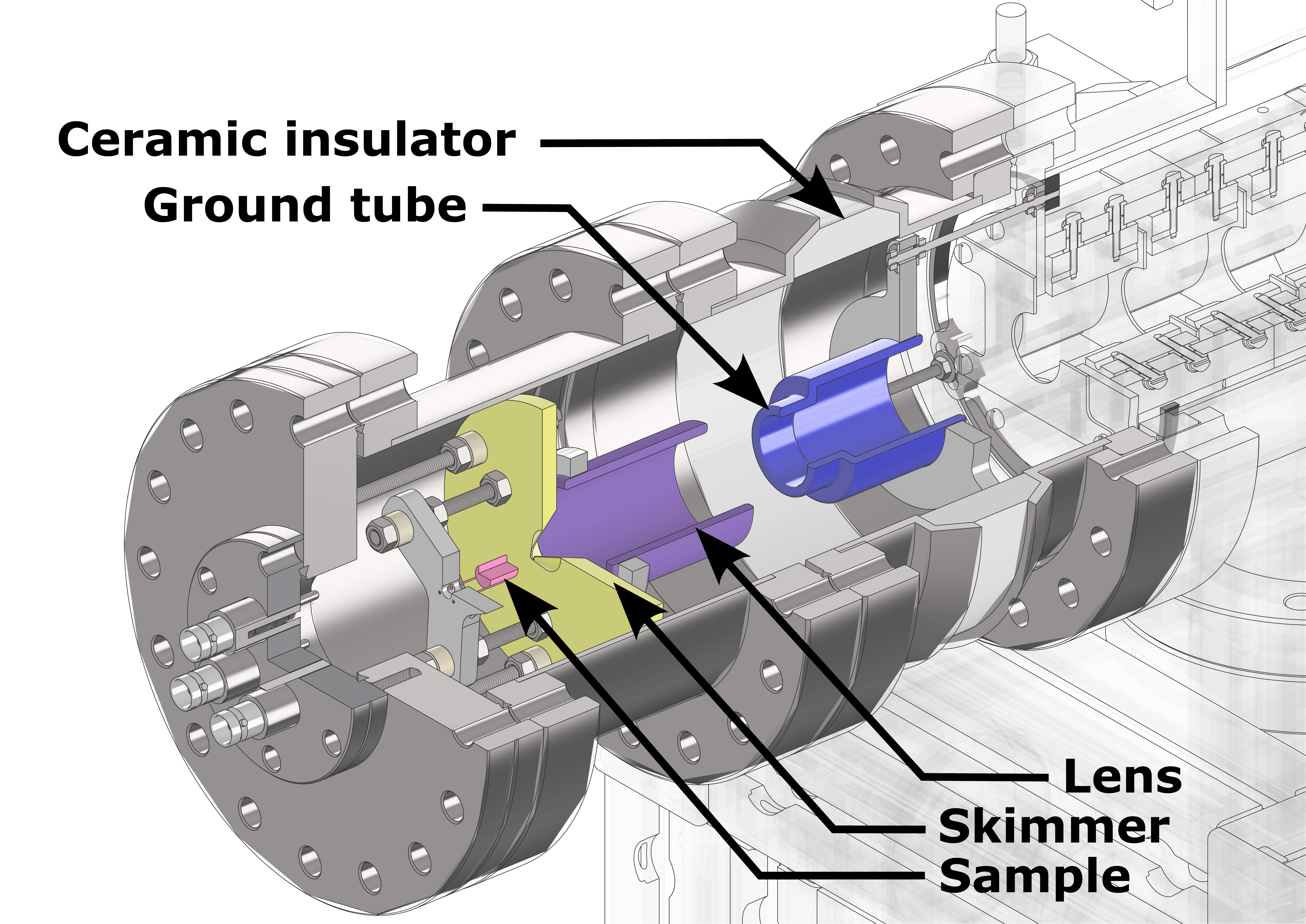}
    \caption{Three-quarter view of the ion source placed at HV and used in the HIBISCUS beamline. The ion source is shown in pink color. The produced ions pass through a skimmer electrode and an electrostatic lens, highlighted with yellow and purple colors, respectively, and get accelerated to ground potential. The hollow tube, in blue color, is at ground potential.\label{fig:source}}
\end{center}
\end{figure}

A surface ionization ion source has been designed to commission HIBISCUS offline. A heatable stable $^{85, 87}$Rb source from cathode.com (model 101139-06) was used to produce ion beams. The design of the source is shown in Fig.~\ref{fig:source}. 

An electrical current $I$ is driven through the source filament on which the sample is deposited, simply creating the ions of rubidium by heating. Typically $\sim1$~pA of ion current is produced by applying $I\sim1.3$~A through the filament.  The body of the source is at the HV potential of the ion source vacuum chamber. Two electrostatic elements, a skimmer and a lens with a $1.5$~mm and $30$~mm diameter aperture, respectively, allow efficient extraction, and a rough shaping and focusing of the beam as the ions pass through before they are finally accelerated to ground potential. 

\subsection{Beam shaping}
\begin{figure}[htb!]
\begin{center}
    \includegraphics[width =1\columnwidth]{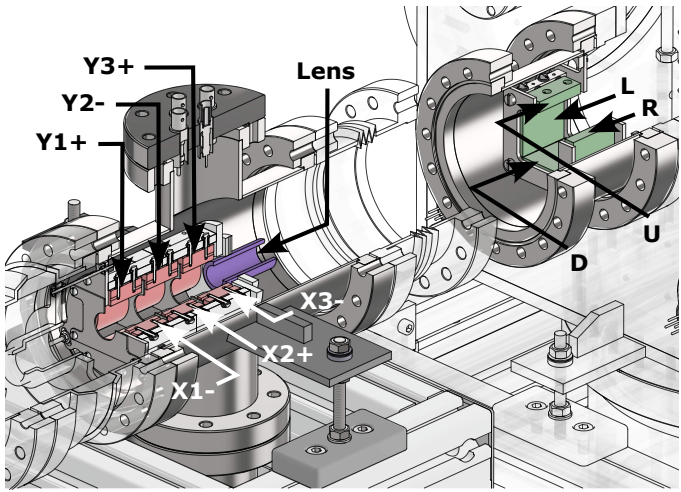}
    \caption{Three-quarter view of the elements focusing and directing the beam in the transverse plane. The triplet electrodes combined with an electrostatic lens are shown in red and purple color, respectively, and the steerers, in green. Voltage polarities of the electrodes are indicated. On the steerers, \textit{L}, \textit{R}, \textit{U}, \textit{D} correspond to left, right, up and down, respectively. The gate valve placed in between the two assemblies is not shown for readability.\label{fig:triplet}}
\end{center}
\end{figure}

As the ion beam needs to be efficiently injected into the RFQCB, its focus and its transverse position have to be optimized. Thus electrostatic elements, laid out in Fig.~\ref{fig:triplet}, are placed prior to the deceleration optics to shape the continuous beam produced by the ion source. A quadrupole triplet first allows to shape the beam by alternatively focusing it the X and Y transverse directions. It consists of a quadrupolar structure segmented in three sets of opposite X and Y cylindrical electrodes with a $20$~mm minimum radial distance between each pair. The electric potentials supplied to the triplet electrodes are chosen so that \textit{X1, Y2, X3} have the same voltage polarity and \textit{Y1, X2, Y3} have the opposite voltage polarity. A 10~mm-diameter electrostatic lens is placed after the triplet and is connected to ground potential.

To compensate for any mechanical misalignments, electrostatic steerers are placed before the injection cell, marked in green in Fig.~\ref{fig:triplet}, and after the extraction cell. The voltage applied on each plate can be independently tuned, allowing the steerers to also provide an additional focusing effect.

One of the injection steerer electrode is used as a beamgate. It is connected to a fast switch whose two-state voltages are set to either fully transmit or deflect the ion beam. Adjusting the switching cycle allows to control the beam intensity injected in the RFQCB.

\subsection{Deceleration optics}
\begin{figure}[htb!]
\begin{center}
\includegraphics[width =\columnwidth]{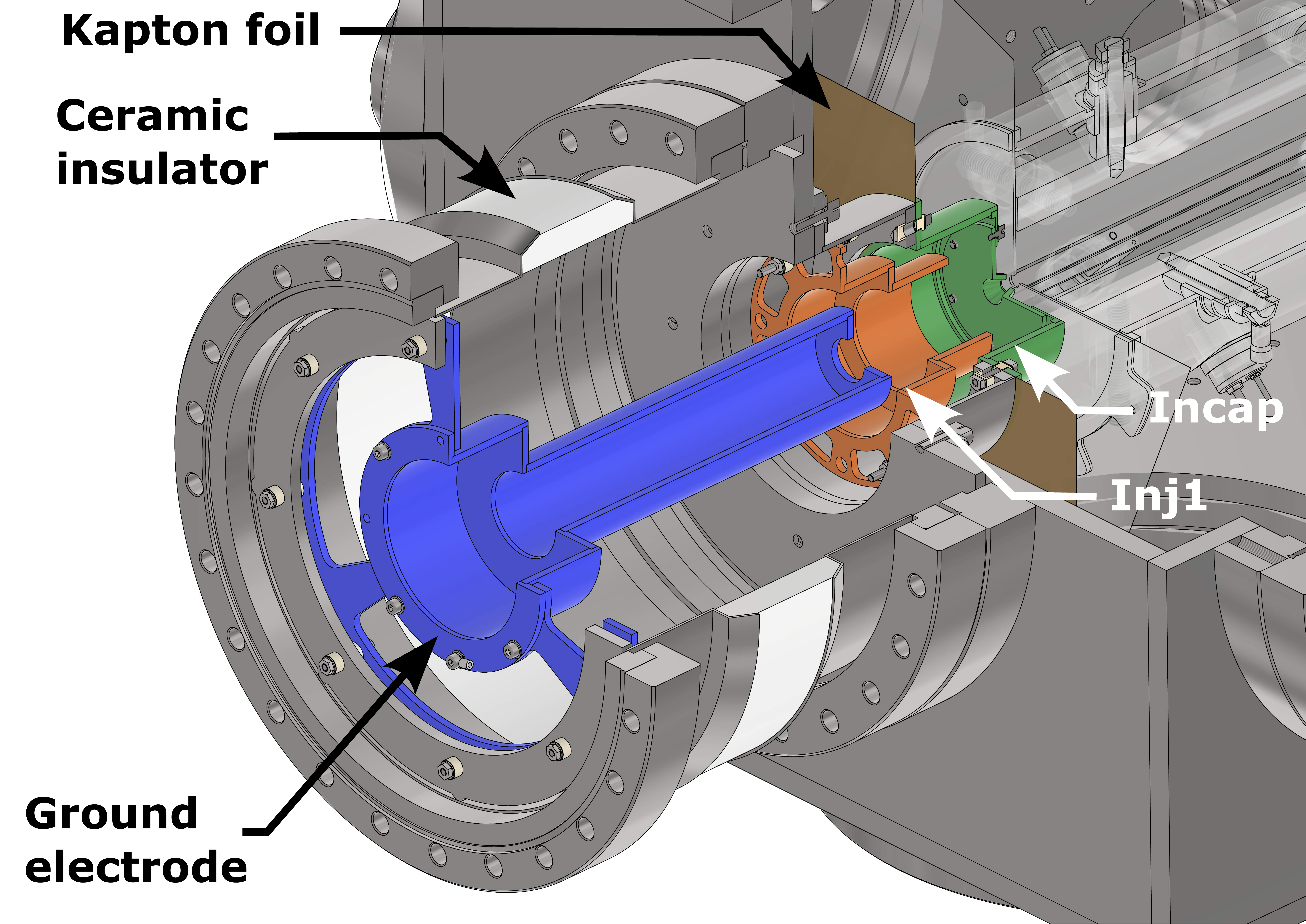}
\caption{Three-quarter view of the deceleration optics of HIBISCUS. The ions enter from the left through the ground electrode (in blue) and pass through two deceleration electrodes (\textit{inj1}, \textit{incap} highlighted in orange and green colors, respectively). A ceramic insulator separates the injection beamline at ground potential from the RFQCB chamber at HV. Kapton foil is used to electrically isolate the \textit{incap} electrode from the main vacuum chamber. The injection side of the RFQ cooler stage is shown in transparency.\label{fig:inj}}
\end{center}
\end{figure}

The ions need to be (electrostatically) decelerated as much as possible before they enter the RFQ structure. In practice, the ions need to have some tens of eV left for an efficient injection. The remaining energy is then cooled in collisions with the buffer gas. The setup is designed to efficiently handle ion beams with kinetic energy up to 30~keV. After a long ground electrode shielding the ions from any spurious potential on the ceramic part of the HV insulator, there are two cylindrical electrodes, \textit{inj1} and \textit{incap}, visible in Fig.~\ref{fig:inj}, to tune the deceleration gradient for maximizing the transmission of ions.

The 5~mm-diameter aperture of the \textit{incap} acts as a pumping barrier between the injection cell and the main chamber housing the RFQCB enclosure.

\subsection{RFQ cooler and buncher} 
\subsubsection{Ion cooling}
\begin{figure}[htb!]
\includegraphics[width =\columnwidth]{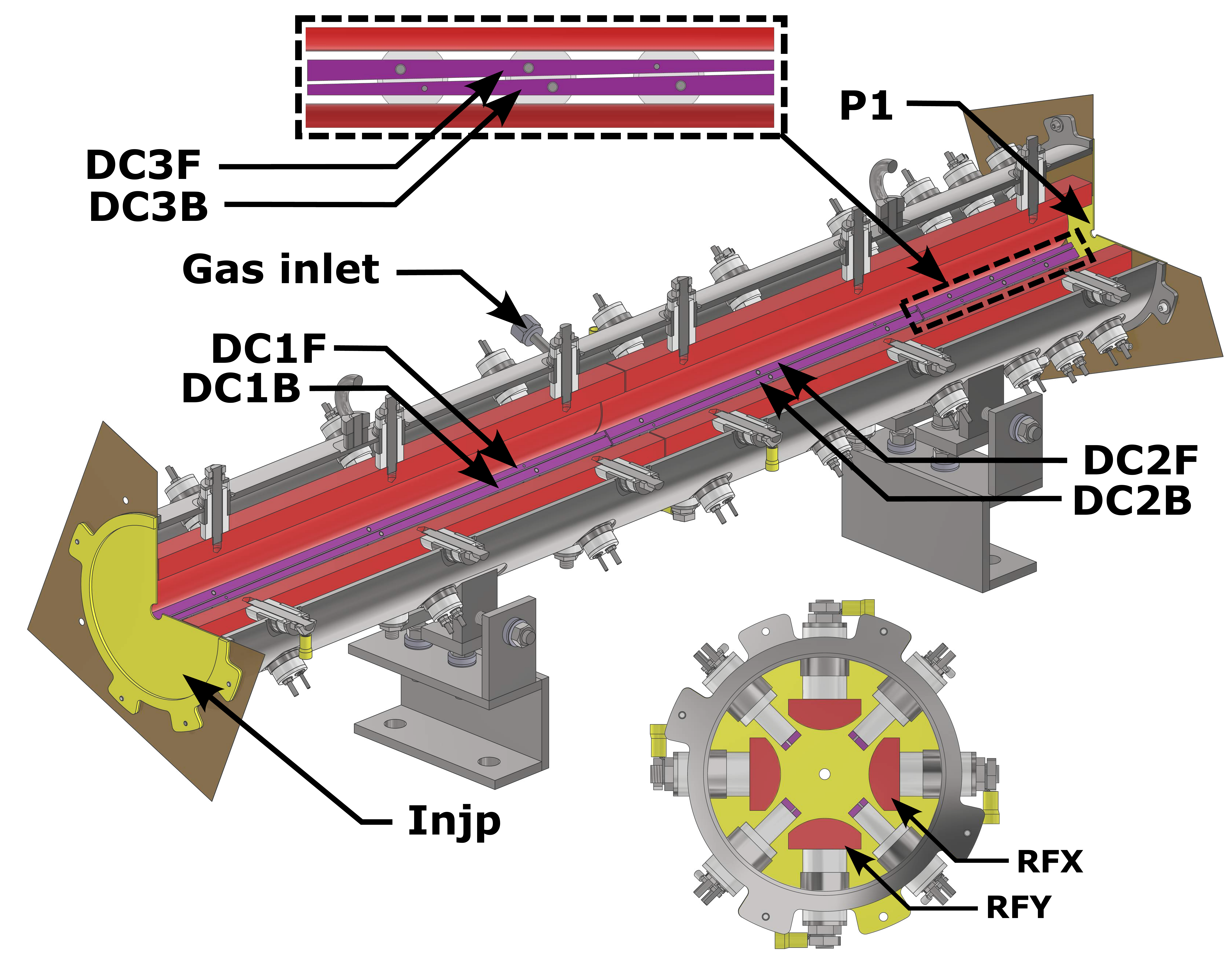}
\caption{Three-quarter view of the RFQ cooling section of HIBISCUS together with a close-up look on its transverse cross-section. Once the decelerated ions enter the enclosure through \textit{injp}, highlighted in yellow color, they are cooled in the volume filled with buffer gas and kept radially confined on axis using a RF quadrupole structure, in red color. The ions are guided towards the exit on the right with a smooth potential gradient applied on DC wedges in purple color. The ions are further extracted to the RFQ bunching sections through \textit{P1} in yellow color, or confined in the \textit{DC3B/F} region. A close-up look on \textit{DC3B/F} shows the shape of the pair of wedges placed in between the RF rods. \label{fig:cool}}
\end{figure}

After the ions pass through the deceleration electrodes, they enter the RFQ cooler volume through a plate with a 10~mm diameter aperture (\textit{injp} in Fig.~\ref{fig:cool}).

The enclosure houses a set of four rods in quadrupolar configuration on which a time dependant electric field is applied to yield a transverse pseudopotential. The enclosure is filled at $\sim0.1$~mbar of pressure with ultra pure helium gas ($99.9999\%$, grade 6.0) fed through a 1/4" gas line. As the ions are injected, they remain confined radially by the RF quadrupolar field while thermalizing on the beam axis via multiple elastic collisions with helium atoms \cite{DEHMELT196853,KELLERBAUER2001276}. The stability of the transverse confinement by the RF field is dictated by the Mathieu equations and parameters $a$ and $q$ \cite{mathieu1868memoire,meixner2006mathieu}, defined respectively as,
\begin{equation}
a=\frac{4QU_{DC}}{m\Omega^2r_0^2} \ , \ q=\frac{2QU_{RF}}{m\Omega^2r_0^2}
\end{equation}
where $Q$ is the electric charge of the ion, $U_{DC}$ is the amplitude of the DC quadrupole field, $U_{RF}$ is the RF signal amplitude (in peak-to-peak voltage, V$_\text{pp}$) applied on the rods, $m$ is the mass of the ion, $r_0$ is the minimum half distance across every pair of opposite rods and $\Omega$ is the angular frequency. Hence, stability of the ion motion depends on the mass-over-charge ratio of the ions and electrical and geometrical constraints. In absence of a DC quadrupole component it follows that $a=0$, and the ion motion is stable when $0<q<0.908$ \cite{paul1955elektrische}.

With their axial and transverse energies being minimized in the cooling process, a small and smooth electrostatic gradient is also applied to guide the ions towards the exit of the RFQ cooler. This axial electrostatic component modifies the equations of the ions' motion, giving a non-zero $a_{eff}$:
\begin{equation}
   a_{eff}=\frac{4QU_{DC_{eff}}}{m\Omega^2r_0^2}
\end{equation}
where $U_{DC_{eff}}$ is proportional to the DC component axially applied and depends on the geometry of the RFQ cooler section \cite{drewsen2000harmonic}. Stable transfer of the ions is ensured by having $U_{DC_{eff}}<0$, of typically a few volts, which is applied using the DC wedge electrodes (see Fig.~\ref{fig:cool}).

\subsubsection{Ion bunching}
The RFQ bunching section differs from the RFQ cooling section on the DC voltage application. The quadrupolar structure is segmented to allow application of DC voltage on top of the radially-confining time-dependent electric potential (see Fig.~\ref{fig:bunch}). These allow, together with the electrostatic plates placed on the enclosures' ends, to shape a potential well to axially confine the ions. The stability of the trapped ions' motion in a potential well is characterized by ($a,q$) parameters in a similar manner as in the RFQ cooler section. Suitable combination of the Mathieu parameters ensures that the ion's trajectory is stable throughout the sections.

After the ions are properly transferred from the main cooling stage to the subsequent bunching sections, and gathered in the potential well, they can be extracted as a bunch from the enclosure, through a plate with a narrow aperture.

\subsubsection{Technical design}
\begin{figure}[htb!]
\includegraphics[width =\columnwidth]{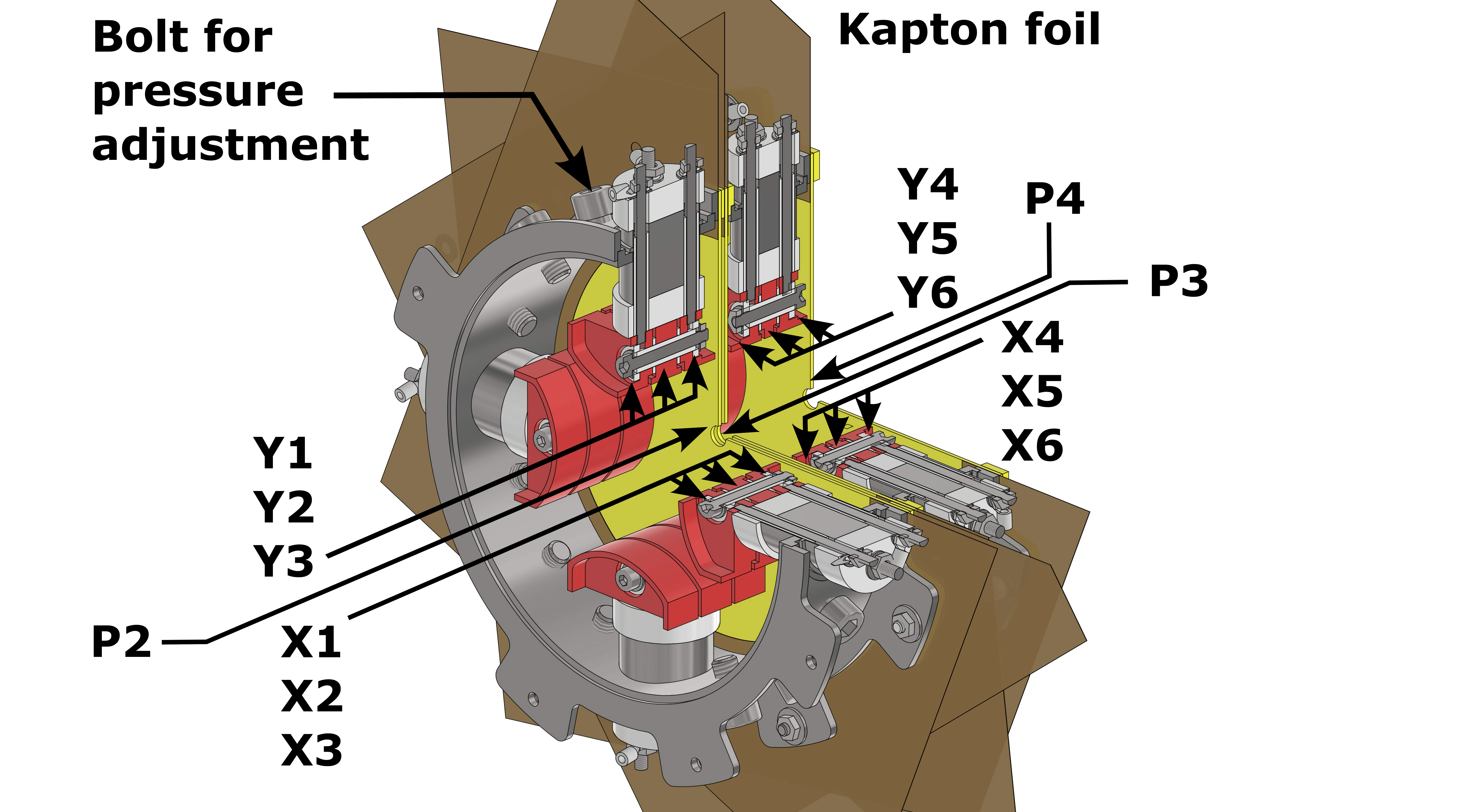}
\caption{Three-quarter view of the RFQ buncher sections of HIBISCUS. Once the ions exit the RFQ cooler section and enter the first buncher enclosure through the \textit{P1} plate (see Fig.~\ref{fig:cool}), they are confined axially and radially using a quadrupole structure, shown in red color. The ions are then transferred to the second buncher enclosure through the plates \textit{P2} and \textit{P3}, where their motion remains controlled by a similar structure. They are finally extracted towards the acceleration optics through \textit{P4}. Kapton foil is used to insulate the \textit{P2}, \textit{P3} and \textit{P4} electrodes, highlighted in yellow color, from the other RFQ buncher electrodes. \label{fig:bunch}}
\end{figure}

The minimum distance across every pair of opposite rods is $2r_0=40$~mm in all stages of the RFQ structure, see Fig.~\ref{fig:cool} and Fig.~\ref{fig:bunch}. High-amplitude RF field, ranging from $U_{RF}=10-1300$~V$_\text{pp}$ at $f=300-600$~kHz is applied throughout the device, tuned so that facing rods have the same phase while adjacent ones are in opposite phase. 

The RFQ cooler stage consists of four RF cylindrical rods (opposite \textit{RF X}s and \textit{RF Y}s), each constructed of two 400-mm long rods for a total length of 800~mm. For DC gradient application, four pairs of electrostatic wedges are mounted between the RF rods and are segmented in three progressively shorter stages towards the extraction side, from 400~mm long, to 260~mm long and finally 130~mm long (\textit{DC1B/F}, \textit{DC2B/F} and \textit{DC3B/F}, respectively, as laid out in Fig.~\ref{fig:cool}). For each pair, the wedge 'front' facing the beam, i.e. oriented with the larger end first, is labeled with the letter \textit{F}. The wedge 'back' facing the beam, i.e. oriented with the smaller end first, is labeled with the letter \textit{B}. These allow a finer application of the DC gradient towards the end of the section. This arrangement also allows to form a trapping potential region at the far end of the RFQ cooler stage when the potentials on \textit{DC3B/F} are set to be lower than the neighboring \textit{DC2B/F} and \textit{P1} potentials.  The plate with a narrow aperture \textit{P1}, can then have its voltage fast-switched to block the ions or let them be released. 

The RFQ buncher stage is separated into two enclosures, each containing three segments of RF quadrupole rods. This allows to shape the potential on the beam axis in each bunching sections. Among these six segments, opposite \textit{X1-6} and \textit{Y1-6}, five are 10~mm long, and one is 3~mm long (namely opposite \textit{X5, Y5}), as shown in Fig.~\ref{fig:bunch}. The DC potential applied on each quadrupole segment generates either an axial trapping well or a gradient depending on the voltage configuration. These two sections are enclosed by plates, \textit{P1, P2, P3} and \textit{P4} and the voltages applied on those can be fast-switched for transfer and extraction of the ion sample. The enclosures for the RFQ cooler and buncher stages, are not completely vacuum tight. As helium is fed through the gas line to the RFQ cooler enclosure, the only paths for the gas to exit is through \textit{injp} and \textit{P1} apertures to the main RFQCB vacuum chamber volume and the RFQ buncher sections, respectively. To control the amount of gas present in the RFQ buncher stage, bolts situated all around the enclosures can be removed, allowing to fine tune the pressure by letting more of the gas to be pumped away. All the diaphragm electrodes \textit{P1-P4} have a 5~mm-diameter aperture.

\subsection{Acceleration optics}
\begin{figure}[htb!]
\begin{center}
\includegraphics[width =\columnwidth]{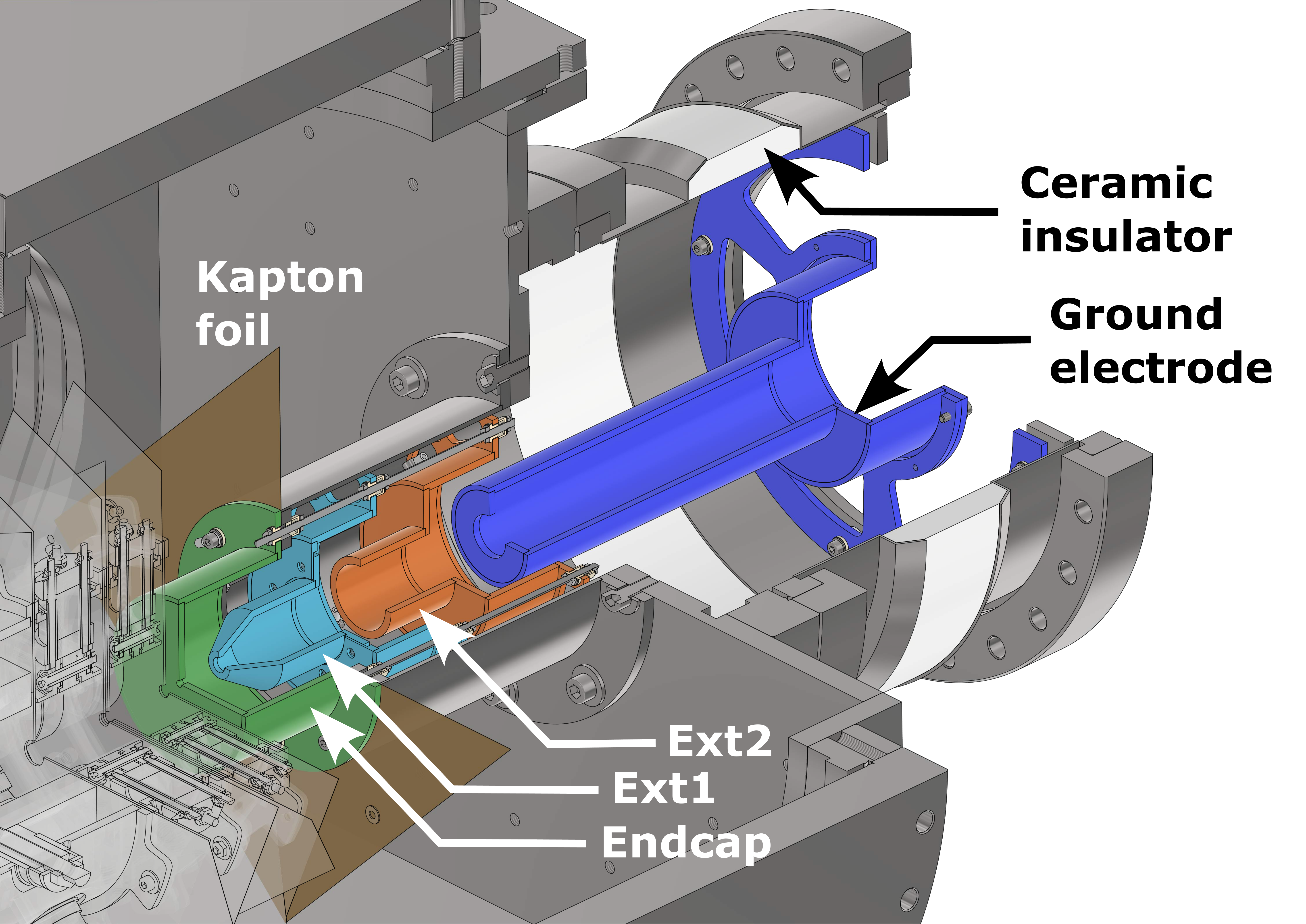}
\caption{Three-quarter view of the acceleration optics of HIBISCUS. After extraction from the RFQ buncher stage via \textit{P4}, the ions pass through three electrodes (\textit{endcap}, \textit{ext1}, \textit{ext2}, highlighted in green, turquoise and orange colors, respectively), and finally exit to the right through the ground electrode, in blue color. A ceramic insulator separates the ground section of the beamline from the RFQCB chamber at HV. Kapton foil is used to float the \textit{endcap} electrode from the RFQCB chamber. The RFQ buncher stage is shown in transparency.\label{fig:ext}}
\end{center}
\end{figure}

The cooled and bunched ions are accelerated to their initial incident energy and transported towards the downstream experimental area. Similarly to the deceleration system, the extraction is composed of three cylindrical electrostatic elements, namely \textit{endcap}, \textit{ext1}, \textit{ext2}, to tune the accelerating field, and a ground electrode, visible in Fig.~\ref{fig:ext}. The small $3$~mm diameter aperture of the \textit{endcap} acts as a pumping barrier between the RFQCB chamber and the extraction cell. 

\section{Operation}
\subsection{Gas and vacuum system}
\begin{figure}[htb!]
\includegraphics[width =0.95\columnwidth]{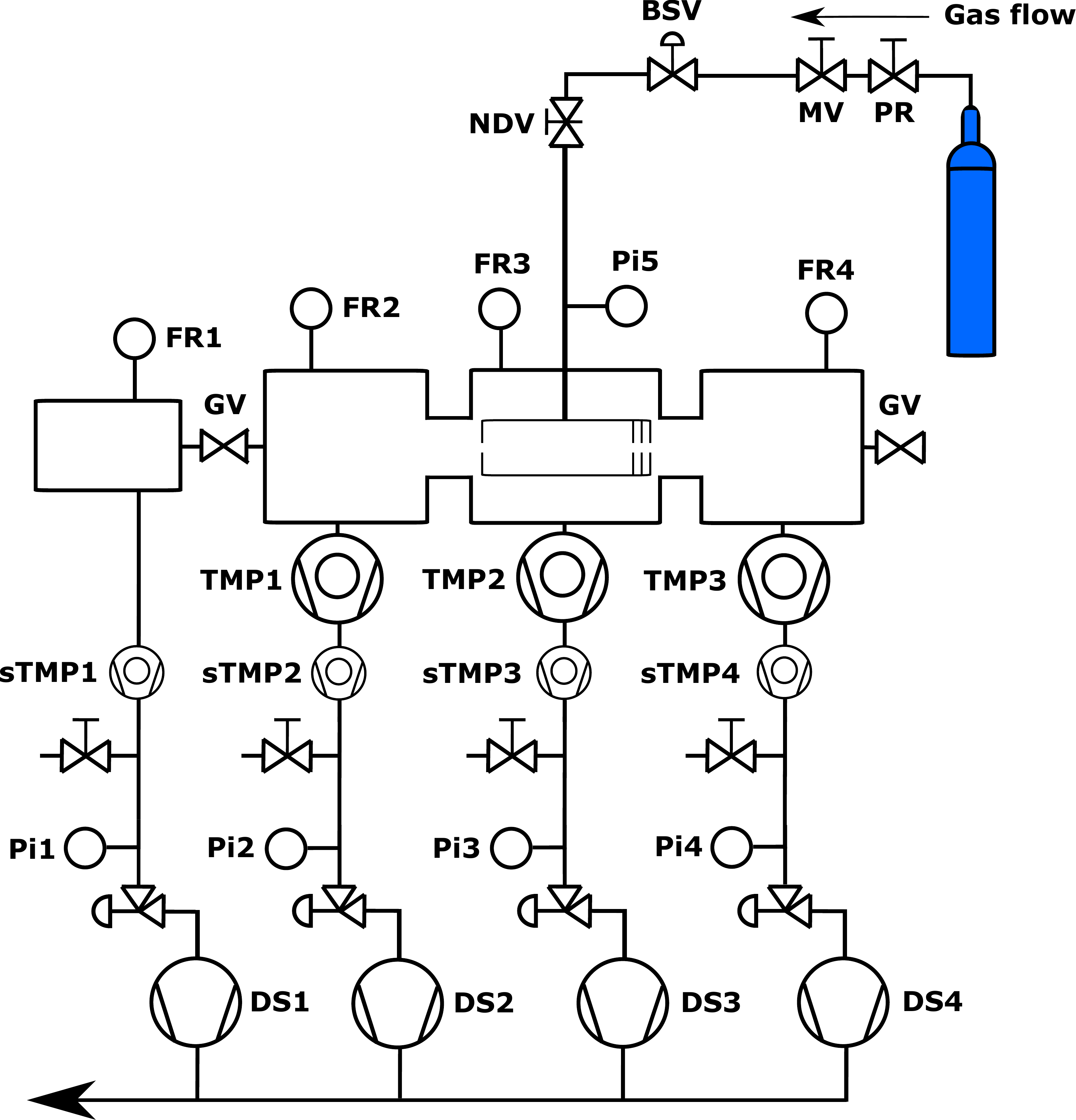}
\caption{Layout of the vacuum system of HIBISCUS\label{fig:gas}}
\end{figure}

A schematic of the vacuum system of HIBISCUS is shown in Fig.~\ref{fig:gas}. The main section is enclosed between the gate valves (\textit{GT}) and pumped through three pumping stations, each of them being a combination of a main water-cooled turbomolecular pump STP-iS2207 (2200~L/s), an air-cooled turbo molecular "pre-pump" nEXT85H (75~L/s) to increase the compression ratio, and a dry scroll pump nXDS6i (\textit{TMP}, \textit{sTMP}, \textit{DS}, respectively). A fourth station composed of a \textit{sTMP} and a \textit{DS} pumps is used for pumping both the quadrupole triplet and the ion source chambers. 

Helium gas is fed from a grade 6.0 (99.9999$\%$ purity) bottle. A pressure reducer (\textit{PR}) placed at the helium bottle outlet permits the regulate the pressure to $\sim$2.2~bar of absolute pressure in the 1/4"-diameter gas line. At minimum, $\sim$2~bar pressure is needed to avoid HV discharges inside the gas line ceramic section isolating the HV area from the ground potential. A manual on-off valve (\textit{MV}) in the gas line on the ground side and a pneumatic bellow-sealed valve (\textit{BSV}) on the HV side allow to shut off the gas flow when not needed. Further away, closer to the main chamber, a manual sapphire-sealed variable leak valve (\textit{NDV}) is used to fine controls the gas pressure inside the RFQCB structure with a maximum inlet flow of $6.0$~L/min. The gas load in the gasline and the pressure in the various parts of the RFQCB structure have been simulated using COMSOL Multiphysics \cite{multiphysics1998introduction} with the assumption of an incompressible laminar flow in all regions. The simulations suggest the gas flow needs to be in the 0.01~mbar l/s range in the gas line to reach a few 10$^{-2}$~mbar of helium pressure in the RFQ cooler cavity, and in the 1~mbar l/s range to reach a few 10$^{-1}$~mbar. The base vacuum in HIBISCUS's chambers without gas feeding reaches $2\times 10^{-8}$~mbar, and $5\times 10^{-4}$~mbar in the region of the gas inlet to the RFQ cooler enclosure. According to the simulations, this corresponds to a residual gas pressure of $\sim10^{-5}$~mbar in the RFQCB enclosure, mainly originating from atomic and molecular desorption from the surfaces inside. Operating pressure levels in different sections of the beamline considering a typical helium gas feeding pressure are summarized in Tab.~\ref{tab:oppressures}. 

Manual venting valves are present on the side of the \textit{sTMP} turbomolecular pumps and the dry scroll pumps can be isolated from the system by closing off manual angle valves  installed on top of the pump. Full range pressure gauges PKR251 are placed on the side of the main vacuum chamber (\textit{FR3}) as well as on the injection (\textit{FR2}), extraction (\textit{FR4}) cells, and on the ion source and quadrupole triplet (\textit{FR1}) chambers. Pirani gauges TPR280 indicate the pressure in the RFQCB gas line (\textit{Pi5}) and monitor the dry scroll pump pressures (\textit{Pi1}, \textit{Pi2}, \textit{Pi3} and \textit{Pi4}).

\begin{table}
 
    \centering
    \begin{tabular}{ll}
       Section & Pressure (mbar)\\ \hline \hline
       Quadrupole triplet & $2.6\times 10^{-7}$\\
       Injection cell & $3.1\times 10^{-7}$\\
       RFQCB chamber & $1.7\times 10^{-5}$\\
       Extraction cell & $3.3\times 10^{-7}$\\
       Dry scroll pump lines & $10^{-2}-10^{-1}$
    \end{tabular}
    \caption{Typical vacuum levels read from the vacuum gauges in the different sections of HIBISCUS when all pumps are running at their nominal capacity. Unfortunately, the pumping station in the injection beamline became out of service during the commissioning, increasing the pressure levels in the quadrupole triplet section and injection cell to the low 10$^{-5}$~mbar range. Though it does not affect the properties of the extracted ions from the RFQCB, it decreases the efficiency of their injection into the structure.}
    \label{tab:oppressures}
\footnotetext[1]{}
\end{table}

\subsection{Modes of ion extraction \label{modes}}
\begin{figure}[htb!]
\includegraphics[width =1\columnwidth]{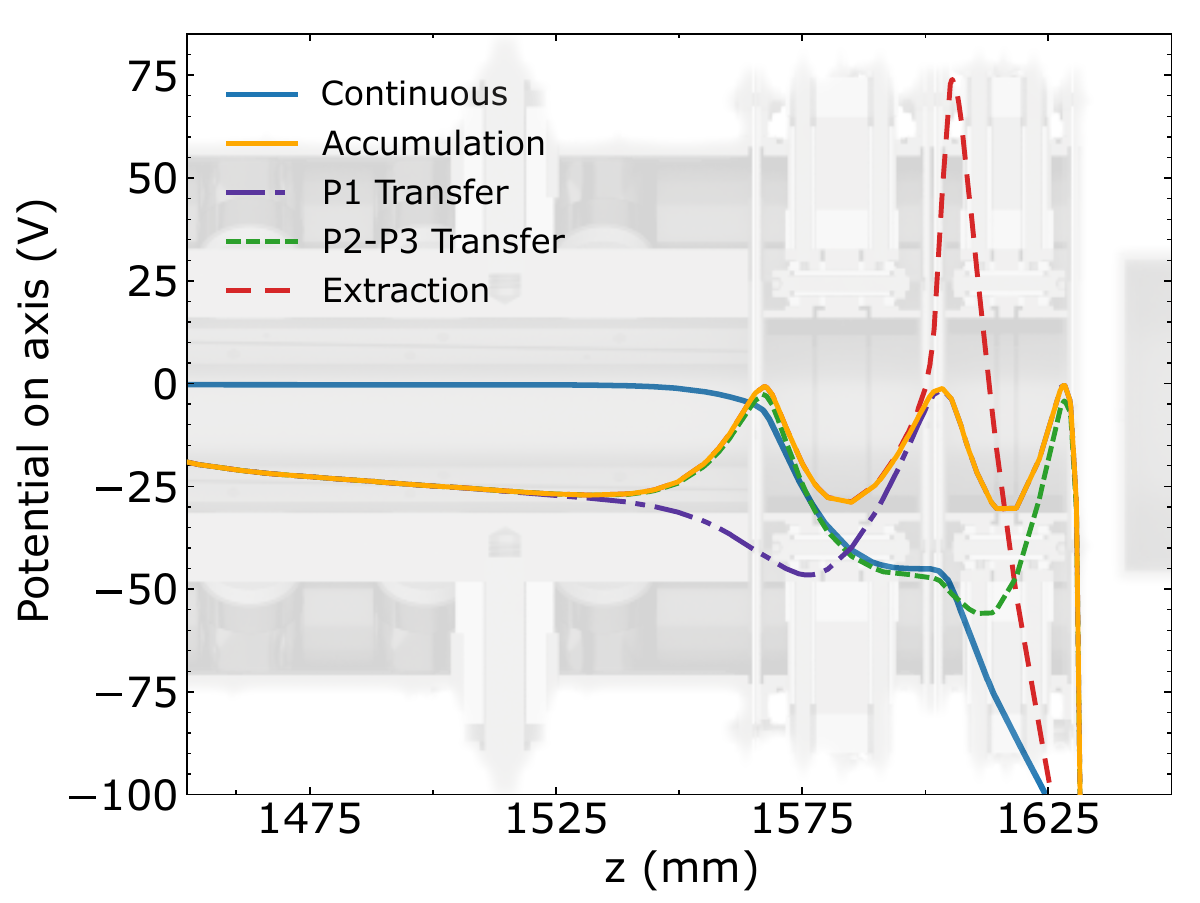}
\caption{Typical DC potentials on the beam axis of HIBISCUS for different modes of extraction. The blue curve corresponds to the continuous extraction mode. The orange, purple, green and red curves shows the different steps of the bunching process, here in the case of a "fast" extraction. As the ions are thermalized the same way in the RFQ cooler, the potential evolution is the same in this section for all modes. These effective potentials on the beam axis of HIBISCUS are obtained using COMSOL Multiphysics \cite{multiphysics1998introduction}.  \label{fig:pot}}
\end{figure}

Depending on the beam requirements needed for the downstream experimental setups, HIBISCUS can be operated with three different modes of extraction. After being cooled, the ions can be either (i) extracted continuously, (ii) as bunches with a minimized energy spread, or (iii) with a best-reduced temporal width at the time focus. In continuous mode, the plate electrodes with narrow apertures \textit{P1, P2, P3} and \textit{P4} are set to their default "low" voltage state and the DC component applied on the segments of the RFQ buncher stage are set to efficiently transfer the ions through the sections. How the electric potential on the beam axis looks in this mode is shown in blue in Fig.~\ref{fig:pot}.

When the ions are extracted as bunches, trapping wells to gather the ions in both RFQ buncher sections are shaped by adequately setting the DC voltages applied on the quadrupole segments and, \textit{P1, P2, P3} and \textit{P4} are switched sequentially to transport and extract the formed bunch using fast-switches. Each full cooling and bunching cycle in the order of a few milliseconds long, giving a switching rate of $\textit{P1-P4}$ of typically a few hundred Hz.

The first step, in orange in Fig.~\ref{fig:pot}, consists of accumulating the ions in a potential well at the far-end of the RFQ cooler section by having \textit{P1} in its "high" state. The ions are there gathered and kept cooled before being transferred to the first RFQ buncher section. To achieve the transfer, \textit{P2} is kept in its "high" state, while \textit{P1} is switched to its "low" state, allowing the ions to pass through the narrow aperture. The corresponding potential evolution on the beam axis is plotted in purple in Fig.~\ref{fig:pot}. The ions are then shortly captured in the first RFQ buncher section by switching back \textit{P1} to its "high" state, before finally being transferred to the second RFQ buncher section. This is made by keeping \textit{P1} and \textit{P4} in their "high" states while \textit{P2} and \textit{P3}, the plates separating the two bunching enclosures, are in their "low" states. The corresponding potential evolution on the beam axis is plotted in green in Fig.~\ref{fig:pot}. \textit{P2} is switched back to its "high" state after the transfer to capture the ion sample. Meanwhile, a second bunch can be loaded in the far-end of the RFQ cooler section, paralleling the process. After some time, the bunch is finally extracted from the second RFQ buncher section by having \textit{P3} in its "high" state and \textit{P4} in its "low" state, and is further accelerated. 

Considering the axial oscillation of the captured ions, they are released with a certain distribution of kinetic energy when the potential is fast-switching from a trapping well to a linear extraction gradient. The strength of this extraction field specifically affects the properties of the ion bunch in terms of time and energy spreads. With strong fields, i.e. steep potential gradients, the ions furthest away from the exit side will gain more energy upon extraction compared to the ones nearest to the exit side. The temporal width of the bunch at the time focus, i.e., when the more energetic ions reach the less energetic ones, is then reduced, at the expense of a wider energy spread. The typical resulting potential slope in this mode is seen in red in Fig.~\ref{fig:pot}. The extraction gradient formed between \textit{P3} in its "high" state and \textit{P4} in its "low" state is around $\sim10$~V/mm. If the aim is on reducing the energy spread of the extracted bunch to its minimum, the velocity gain upon release from the trapping well needs to be small, i.e. small extraction fields are preferred, of typically $\sim0.1-1$~V/mm. This "slow" extraction however gives a longer flight duration for the ion bunch to broaden in time before reaching the detector. The extraction settings then must be adjusted depending on the ion bunch characteristics required by the experimental setups placed after HIBISCUS.

\subsection{Beam diagnostics}
In order to characterize HIBISCUS and to optimize and monitor the ion beam when in operation, a set of diagnostics are housed in the two identical cells, placed on the injection and extraction side of the RFQCB, as labelled in Fig.~\ref{fig:beamline}.

Faraday cups (FC) are used to monitor the beam current from the ion source. With its central metallic plate placed on the beam axis, the FC reads a continuous current signal proportional to the impinged beam intensity. FCs are mainly used to extract transmission efficiency numbers through the experimental setup when HIBISCUS is set to continuous transfer mode. There are four independent segment electrodes besides the central cup, allowing to indirectly map a rough transverse shape of the beam by reading the current on these electrodes separately. A biased suppressor with negative $\sim10-50$~V is also used in front of the FC to push secondary electrons back to the metallic plate surface.

MagneToFs (model "MagneTOF Mini" from etp-ms.com) are used to perform single ion counting. The use of the MagneToF on the extraction side allows to get the time structure of the bunch extracted from the RFQCB thanks to a Time-to-Digital Converter (TDC) Time Tagger (xTDC4 from Cronologic). It records the timestamps of leading and trailing edges of the digital pulses created by detected ion events.

\begin{figure}[htb!]
\begin{center}
\includegraphics[width =1\columnwidth]{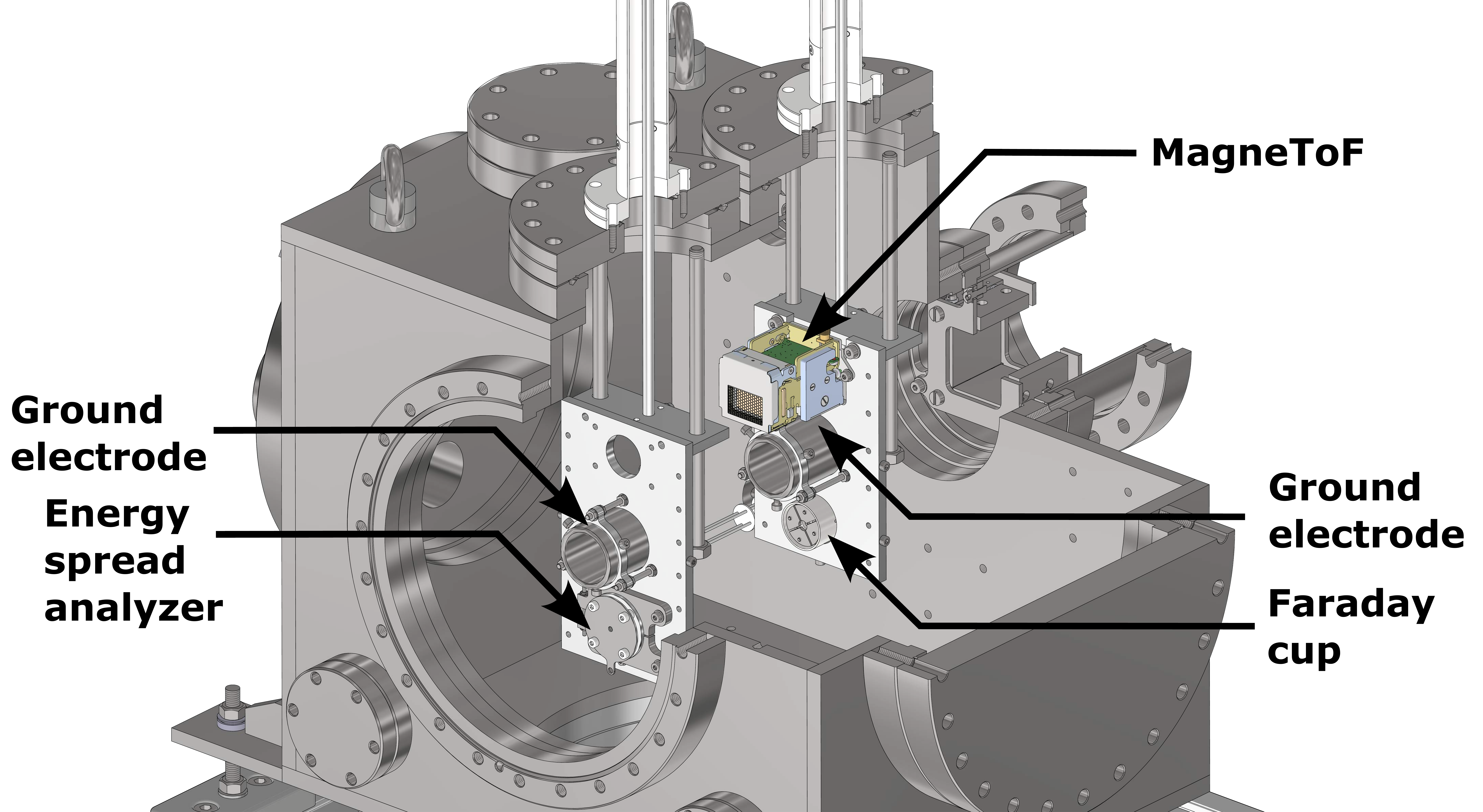}
\caption{Three-quarter view of the extraction cell of HIBISCUS. The chamber houses a MagneToF, a Faraday cup and an energy spread analyzer on vertically driven actuators. The middle position with a hollow tube is used to let the beam pass through when the cell is not in diagnostic mode.\label{fig:diagnostics}}
\end{center}
\end{figure}

The FC and MagneTOF detectors are mounted on a same metallic holder plate, which can be driven vertically in vacuum with an actuator. The plate has two guiding metallic rails that ensure the consistency in the position of the detectors when moved. There are two actuators in both injection and extraction cells.  Each of them is able to switch between three distinct vertical positions: the top and the bottom ones for one of the detectors used in diagnostic mode, and the central one for the electrically grounded hollow tube used in normal operation to let the beam passing through. The actuator layout in the extraction cell is shown in Fig.~\ref{fig:diagnostics}. In one of the positions also figures an energy spread analyzer, currently under development for HIBISCUS.

\subsection{Radio-frequency system}
\begin{figure}[htb!]
\includegraphics[width =1\columnwidth]{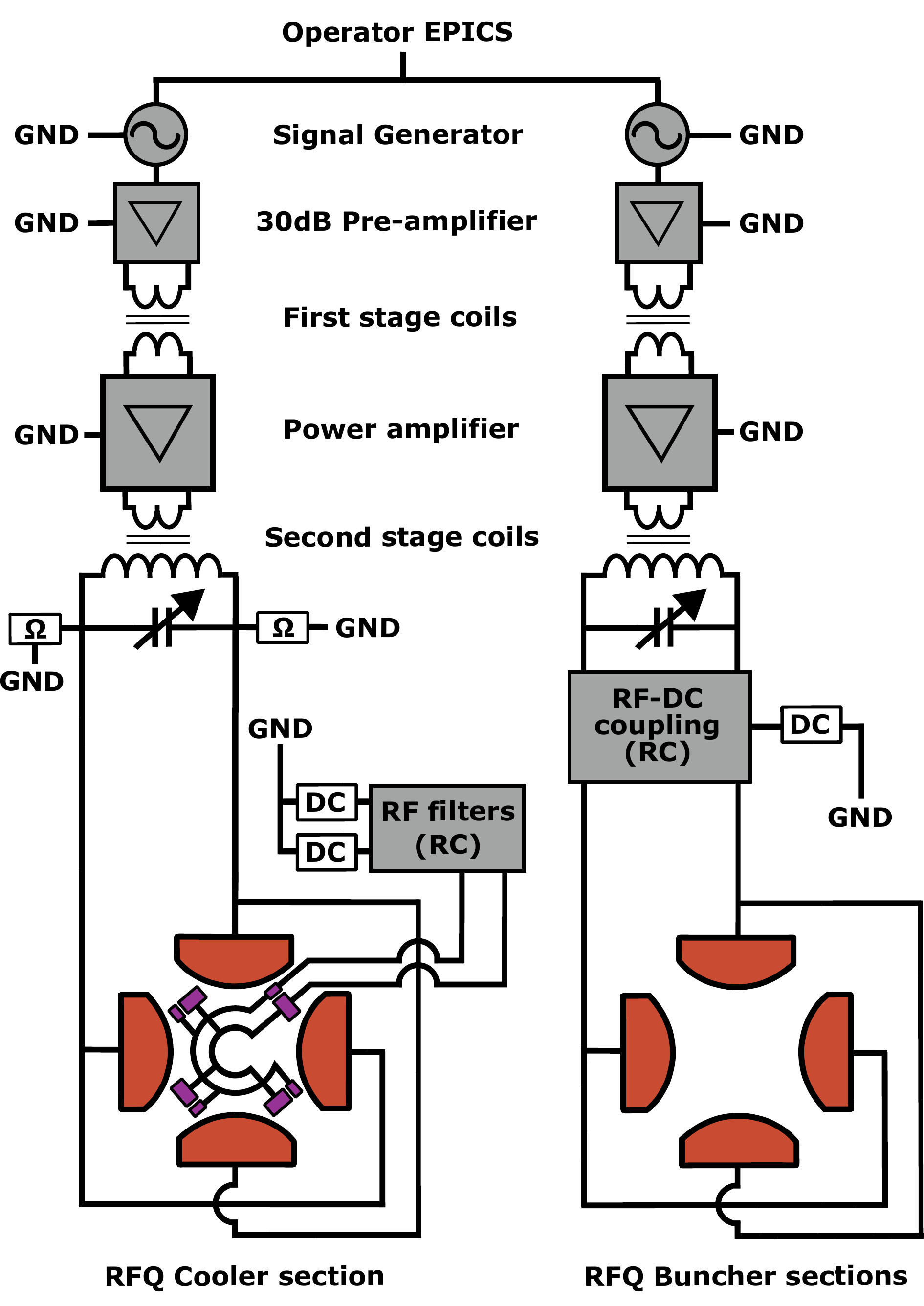}
\caption{Schematics of the RF systems of the RFQ cooler and buncher stages. The RF power circuit and resonant LC circuit are represented for both configurations. The corresponding electrode layout for both sections are also presented for clarity. The RF-DC coupling is identically duplicated for each of the six segments of the RFQ buncher sections.\label{fig:rfscheme}}
\end{figure}

The radio-frequency generation system of HIBISCUS was designed in-house. The RFQ cooler and the buncher stages are powered by their independent driving amplifier and resonant circuits, which provide an oscillating voltage with the same amplitude and frequency to both X and Y pairs of rods and segments, with opposite phase. Having separate amplifier setups for each stage allows to apply different frequency and amplitude to the different stages. Schematics of the equivalent circuits are shown in Fig.~\ref{fig:rfscheme}. Typical RF system parameters and intrinsic capacitive loads of the electrodes of HIBISCUS are summarized in Tab.~\ref{tab:oprf}.

The RF amplification circuit is identical for both systems. It primarily consists of a signal generator (Keysight Trueform 33510B) controlled via EPICS  \cite{dalesio1991epics} outputting a sine wave at the desired frequency and amplitude. This signal is processed through a $30$~dB pre-amplification stage (ZX60-100VHX+) feeding the signal to a broadband power amplifier through a plastic-cored transformer with a winding ratio between the primary and the secondary of 4:1. This ensures the $50$~$\Omega$ impedance of the pre-amplifier output to be matched with the power amplifier input. The broadband power amplifier, based on the use of power MOSFET transistors (IRFP360) in push-pull configuration, can then provide the signal in a frequency band ranging from $300$~kHz to $3$~MHz to the resonant circuit. For safety purposes, temperature sensors decrease the transistor bias voltage if the system heats up after, for instance, being driven out of resonance.

The LC resonant circuit gains and supplies the RF voltage to the RFQ rods and segments with amplitudes up to $1.3$~k$V_\text{pp}$, depending on the circuit parameters. The transformer coils are made with a configuration of NiZn ferrite cores (part number 5967003801 from fair-rite.com) with an initial permeability $\mu_i=40$ and an inductance factor per number of turn squared of $A_L=55^{+35\%}_{-25\%}$~nH. The primary side has one turn. The number of turns on the secondary corresponds to an equivalent $L_\text{cooler}=176$~$\mu$H and $L_\text{buncher}=109$~$\mu$H inductance in the RFQ cooler and buncher stages, respectively. Both X and Y signals are taken through the opposite side of the coil of the same resonant circuit, ensuring that they always remain in opposite phase. The load of the resonant circuit is fully capacitive and has both a fixed intrinsic and a tunable components. The latter is a COMET HV ceramic variable capacitor with a tunable range $150-1500$~pF (withstanding up to 4~kV of $U_{RF}$ at 150~pF and up to 2.4~kV at 1500~pF), added in parallel to the electrodes to adjust the capacitance of the circuits to match the resonant frequency of interest, within the frequency band of the amplifier and depending on the set of transformer coils in use. As specific ion species with their mass-over-charge $m/q$ ratio were used for the commissioning, secondary coils with a high number of turns were made, favoring a higher gain factor at the expense of frequency range tunability with the variable capacitors. In online operation, a compromise needs to be found between the voltage gain factor and the accessible frequency range to cover a wider mass range with a given set of coils.

In the RFQ buncher sections, an additional static voltage component on top of the RF is applied to the segmented electrodes. In order to couple these, an RC low-pass filter (HV 1~M$\Omega$ resistor and 100~nF capacitor to ground, both rated for voltage $>5$~kV) between the DC supply and the electrodes is placed. Similarly, although in the RFQ cooler the RF and DC electrodes are geometrically independent, they remain close enough so the electrostatic wedges are influenced by the RF field through their mutual capacitance. Thus an RC circuit (HV $5$~M$\Omega$ resistor and 100~nF capacitor to ground) is also included on the output of the DC supplies to filter out the picked-up time-varying voltage. 

The LC resonant circuit of the RFQ cooler stage is grounded through an HV $5$~M$\Omega$ resistor but can also be floated to an offset DC voltage if needed. The resonant circuit of the RFQ buncher stage is at the same ground reference as the voltage power supplies providing the static-voltage component to the segments.

\begin{table}
 
    \centering
    \begin{tabular}{ll}
       RF Parameter & Value \\ \hline \hline
        \textit{RFQ Cooler} & \\ \hline
        Capacitance to surroundings & $\sim145$~pF\\
        Mutual X-Y capacitances & $\sim75$~pF\\
        Frequency& 550~kHz\\
        Input amplitude& 0.09~V$_\text{pp}$\\
        Output amplitude& 850~V$_\text{pp}$\\
        &\\
        \textit{RFQ bunchers} & \\ \hline
        Middle capacitance to surroundings & $\sim50$~pF\\
        Outer capacitances to surroundings & $\sim40$~pF\\
        Mutual X-Y capacitances & $\sim25$~pF\\
        Frequency& 550~kHz\\
        Input amplitude & 0.11~V$_\text{pp}$\\
        Output amplitude& 1100~V$_\text{pp}$\\
    \end{tabular}
    \caption{Typical HIBISCUS intrinsic electrode capacitances and operating RF parameters for stable $^{85, 87}$Rb$^+$ ions with 6~keV of energy. Mechanical geometries and electrodes layout are explicitly described throughout Sec.~\ref{exp}. In continuous mode, the RF amplitude in the RFQ buncher stage is set to zero.}
    \label{tab:oprf}
\end{table}

\subsection{Control system \label{control}}
\begin{figure*}[htb!]
\begin{center}
\includegraphics[width =0.89 \textwidth]{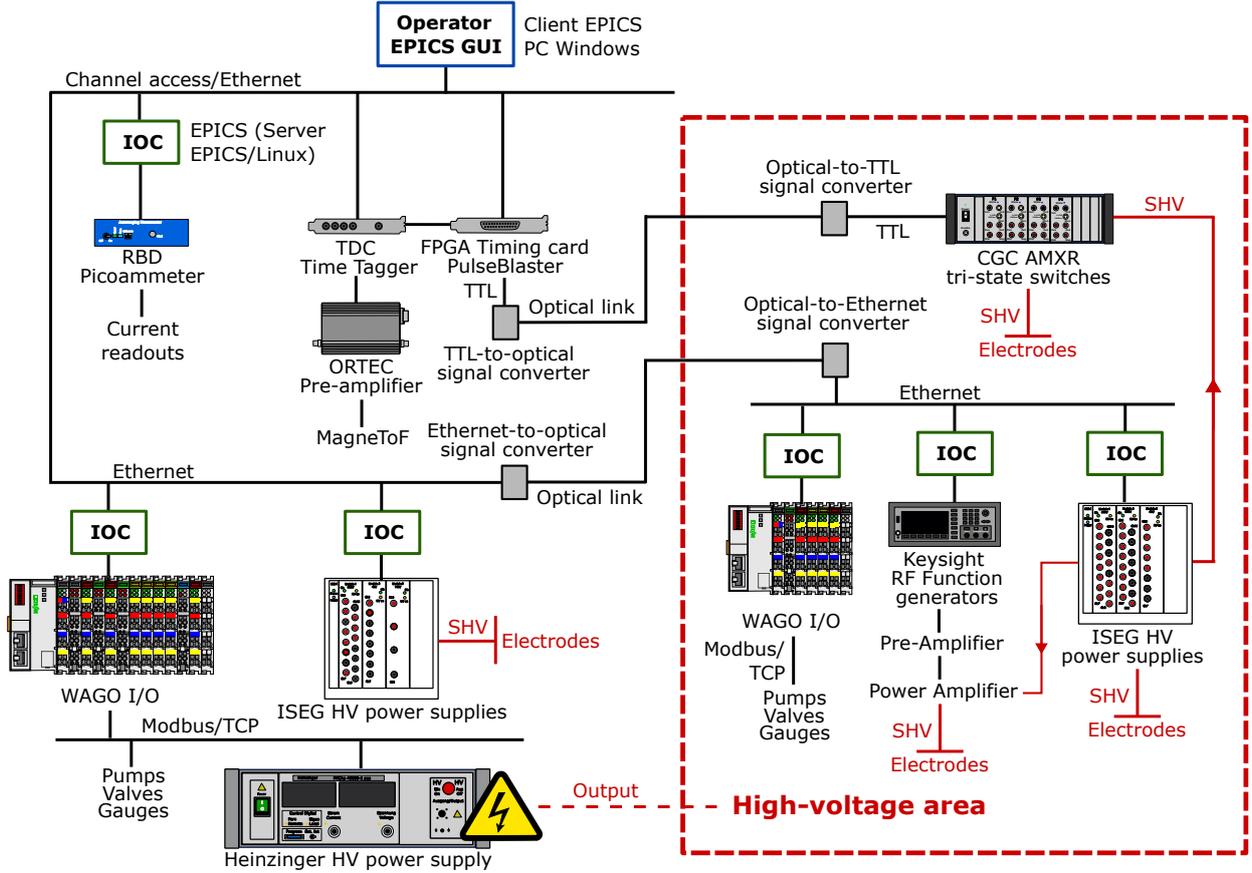}
\caption{Diagram of the control and monitoring system of HIBISCUS\label{fig:control}}
\end{center}
\end{figure*}
An overview of the control and monitoring systems of HIBISCUS is detailed in Fig.~\ref{fig:control}. It is based on the EPICS architecture \cite{dalesio1991epics} with a server running on a Linux virtual machine. The user control of hardware devices is implemented through EPICS-aware graphical user interfaces. Devices use direct digital interfaces when supported by EPICS, with analog and digital I/O modules via the WAGO architecture through MODBUS extension. The only devices implemented on a local control computer are the timing pattern generator (Spincore Pulseblaster) and the Cronologic time tagger. In order to control and monitor the devices placed in the HV area, logic signals are transmitted as optical signals using TTL-to-optical converters. Ethernet is brought to HV using fiber optical cable.

The Heinzinger PNChp-40000-1pos power supply, interfaced with WAGO, provides the HV level of the RFQCB and the ion source up to $40$~kV (with a maximum output current of 1~mA). Here, the HV defines the energy with which the singly charged ions are extracted from the ion source. At FAIR the HV will need to be adjusted to match the incoming 6~keV ion beam energy. The ISEG HV power supplies, directly controlled through an EPICS interface (via CC24), feed continuous adjustable voltages with less than $10$~mV ripple to, the various electrodes of HIBISCUS up to $500$~V (via EHSF505), the injection and extraction optics up to $6$~kV (via EHS8060), and the ion beam shaping electrodes, up to $1$~kV (via EHSF510). ISEGS HV channels are also used to supply the bias for MagneToFs up to $4$~kV (via EHS8240) and the voltage inputs to the CGC AMX500T switch voltage inputs, up to $500$~V. 

The switches have output levels controlled with two TTL-level logic signals. At nominal voltage, the switching time is $<20$~ns, with a fixed propagation delay of $<75$~ns and a minimum pulse duration of $50$~ns. The readbacks of the applied voltages, together with the drawn currents of each channels, are monitored from a dedicated interface. The RF system is being controlled through EPICS with Keysight Trueform 33510B wavefunction generators feeding sine waves with appropriate resonant frequency and amplitude to the amplifiers, up to $20$~MHz and $10$~V$_{pp}$, respectively.

The FPGA card, a PulseBlaster PB24-100-4k-PCIe from Spincore, interfaced locally, is used to control the timings of HIBISCUS. It sends TTL level pulsed signals to the CGC AMXR crate which commands the switching sequence of the plates \textit{P1-P4} in the RFQ buncher sections. The timing card is also used to control the beamgate fast-switch and to give the "start" pulse for the TDC. The timings can be controlled with a 10-ns resolution down to a pulse width of 50~ns.

The detector signal from the MagneToF is first fed to an ORTEC 9326 pre-amplifier and then recorded with a locally interfaced TDC time tagger from Cronologic (model xTDC4) that has a $8$~ps resolution. The TDC has a 218~µs acquisition window for ion hit events following the start signal sent from the FPGA card. If the ion rate exceeds $0.1$~pA of electric current equivalent, the ions are detected with a FC and monitored with a 9103 USB picoammeter from RBD Instruments, interfaced with EPICS.

Finally, the status of the turbomolecular pumps, vacuum gauges, and the control of air pneumatic solenoid valves conducting the various vacuum valves and the HV safety hammer are controlled with WAGO I$/$O modules through EPICS.

\section{Performance in continuous cooling mode}
The tests were done using a 6~keV beam of stable surface-ionized $^{85, 87}$Rb$^+$ ions, cooled and transmitted through the RFQCB while keeping everything constant, i.e. none of the plate voltages applied on \textit{P1}-\textit{P4} were switched. The ion beam current was recorded on the FCs placed in the injection and extraction cells. Optimization of the beam shaping elements, of the injection and extraction optics voltages, and of the DC gradient applied through the RFQ cooler and buncher stages in this mode was also performed. The obtained values were consistently used for these studies. A set of optimum voltages, for ions injected in the RFQCB with $\sim$10~eV of remaining kinetic energy, are reported in Tab.~\ref{tab:opvoltage}.

\begin{table}[htb!]
 
    \centering
    \begin{tabular}{ll}
       Electrode & Voltage value (V) \\ \hline \hline
       \textit{Ion source} & \\ \hline
        HV source & 6010 \\
        Skimmer & 5750\\
        Lens & 5195\\ 
        &\\
        \textit{Steerers} & \\ \hline
        Left/Right & -160/0\\
        Up/Down & 0/130\\ 
        &\\
        \textit{Injection optics} & \\ \hline
        Inj1 & 3965\\
        Incap & 5845\\
        &\\
        \textit{RFQ cooler} &\\ \hline
        Injp & 6005\\
        DC1F/B & 5997.5/5993\\
        DC2F/B & 5996.5/5992\\
        DC3F/B & 5994/5993\\
        &\\
        \textit{RFQ bunchers} & \\ \hline
        1st segment & 5980\\
        2nd segment & 5970\\
        3rd segment & 5980\\
        4th segment & 5990\\
        5th segment & 5985\\
        6th segment & 5990\\
        &\\
        \textit{Switching plates} & \textit{Low}\\ \hline
        P1 & 5995\\
        P2 & 5860\\
        P3 & 5965\\
        P4 & 5730\\ 
        &\\
        \textit{Extraction optics} & \\ \hline
        Endcap & 2300\\
        Ext1 & 5270\\
        Ext2 & 1850\\
    \end{tabular}
    \caption{HIBISCUS operating voltages in continuous cooling mode used for stable $^{85, 87}$Rb$^+$ ions with 6~keV of energy. Electrode labels are explicitly described throughout Sec.~\ref{exp}. It was found that the voltages applied on the quadrupole triplet electrodes were best set at 0~V.}
    \label{tab:opvoltage}
\end{table}

\subsection{Optimization of the Mathieu parameter $q$}
\begin{figure}[htb!!]
\includegraphics[width =\columnwidth]{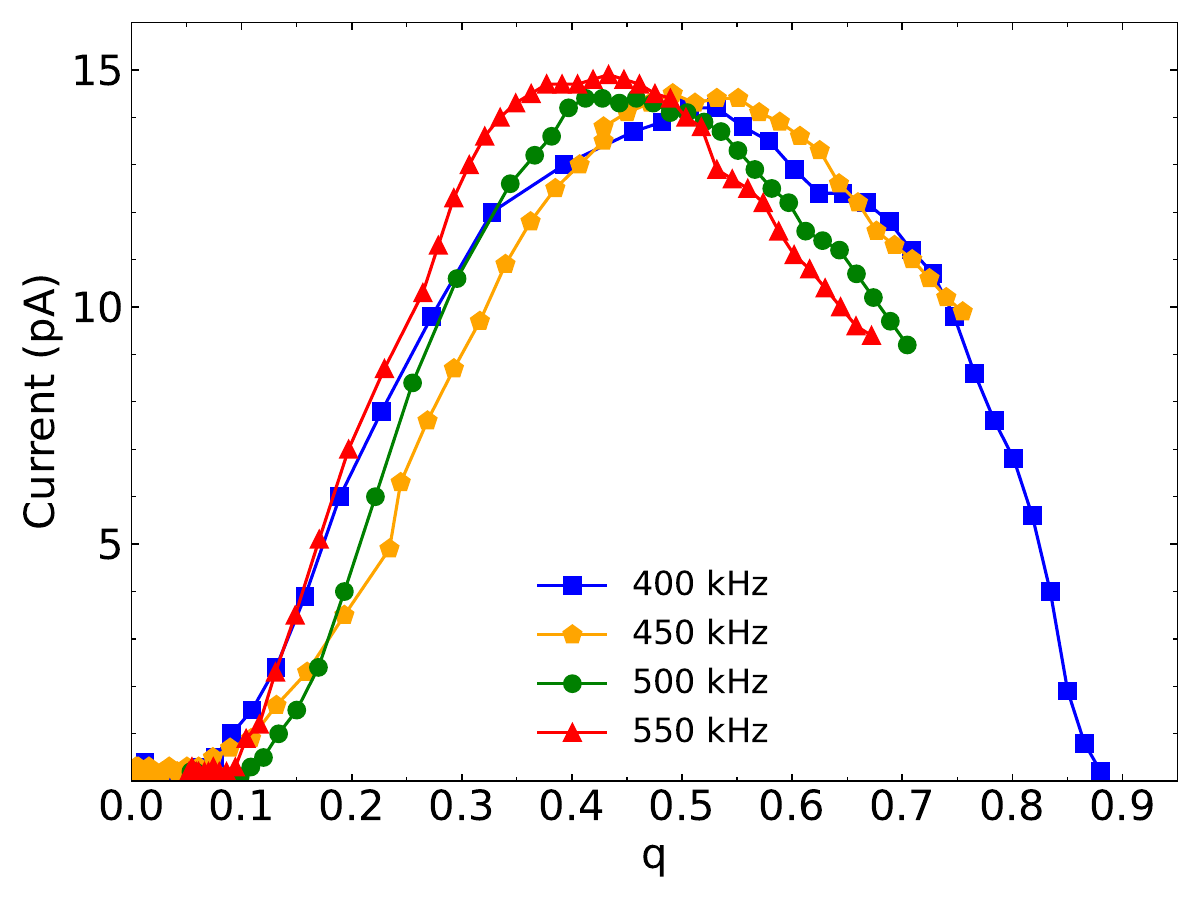}

\caption{Extraction current in continuous cooling mode as a function of the $q$-parameter in the RFQ cooler section for a 6~keV beam of stable $^{85, 87}$Rb$^+$ ions considering different resonant frequencies at a fixed optimum buffer gas pressure and injection energy of the ions. At $f=550$~kHz, $q$ ranging from 0 up to 0.6 corresponds to an RF amplitude ranging from 0 to 1300~$V_\text{pp}$. These measurements were performed with ions injected with 10~eV of remaining kinetic energy and a buffer gas pressure of 0.12~mbar in the RFQ cooler seciton.\label{fig:qparameter}}
\end{figure}

The transverse pseudo-potential provided by the RF field needs to be optimized in order to efficiently radially confine the ions on axis as they are being cooled and transferred through the RFQCB after injection. To do so, the transmitted ion current was recorded while the $q$-parameter of the RF system was varied. More specifically, the amplitude of the RF signal while the frequency $f$ was kept fixed. With HIBISCUS, for $^{85, 87}$Rb$^+$ ions, $q\approx1.4\times10^8 \ U_{RF}/f^2$. The scans, made at $400$~kHz, $450$~kHz, $500$~kHz, and $550$~kHz at fixed buffer gas pressure and injection energy of the incoming ions in the RFQCB, are compiled in Fig.~\ref{fig:qparameter}.

For each considered frequency, the ions are transmitted through the RFQCB when $0<q<0.9$. However for too low $q$-parameter the transmission is poor, as the applied RF field is not strong enough to confine the ions on axis. As the transverse confinement gets more and more efficient with increasing $q$, the transmitted ion current sharply increases, until a plateau is reached, marking a region of optimum $q$-parameter. At higher values, the current starts to drop again until it ultimately falls down to zero for $q>0.908$. This limit is not always reached in Fig.~\ref{fig:qparameter} due to the limitations of the RF system, unable to provide high enough amplitude at higher frequencies.

With the $a$-parameter being negative in a RFQCB, depending on the set $a$, at low $q$ values the working point of operation ($a,q$) may not be within region of stable ion motion. It was found out that the optimal transmission of ions is with $q>0.35$ for the considered frequencies. However, it is clear that the optimum ion transmission interval for the $q$-parameter changes with the applied frequency. For instance, the current drops significantly for $q>0.7$ at $f=400$~kHz, while it already decreases for $q>0.5$ at $f=550$~kHz. This cut-off of the optimum region could be attributed to RF heating effects \cite{DEHMELT196853,PhysRev.170.91}, re-heating the cooled ions and inducing losses. 
At higher frequencies, this effect shows up earlier due to the always higher amplitudes that need to be applied. With $q$ scaling with $U_{RF}/f^2$, a 10$\%$ increase in frequency indeed demands a 21$\%$ increase in RF amplitude for keeping the same set $q$.

Despite the differences in behavior as a function of the $q$-parameter, the maximum output current reached for different frequencies remains comparable. Based on these measurements, it was chosen to operate the RFQ cooler section at $q=0.4$, requiring the application of $U_\text{RF}=850$~V$_{pp}$ at $f=550$~kHz. When HIBISCUS is set to continuous transfer mode, the RF amplitude in the RFQ buncher sections is set to zero.

\subsection{Buffer gas pressure \label{buffer}}
The ions entering the RFQ cooler stage are subsequently cooled by frequent elastic collisions with helium atoms filling the cavity, reducing their remaining kinetic energy as well as the amplitude of their radial motion. The gas pressure is typically chosen to be on the order of 0.1~mbar, giving the injected ions a mean free path of a few mm.

\begin{figure}[htb!!]
\includegraphics[width =\columnwidth]{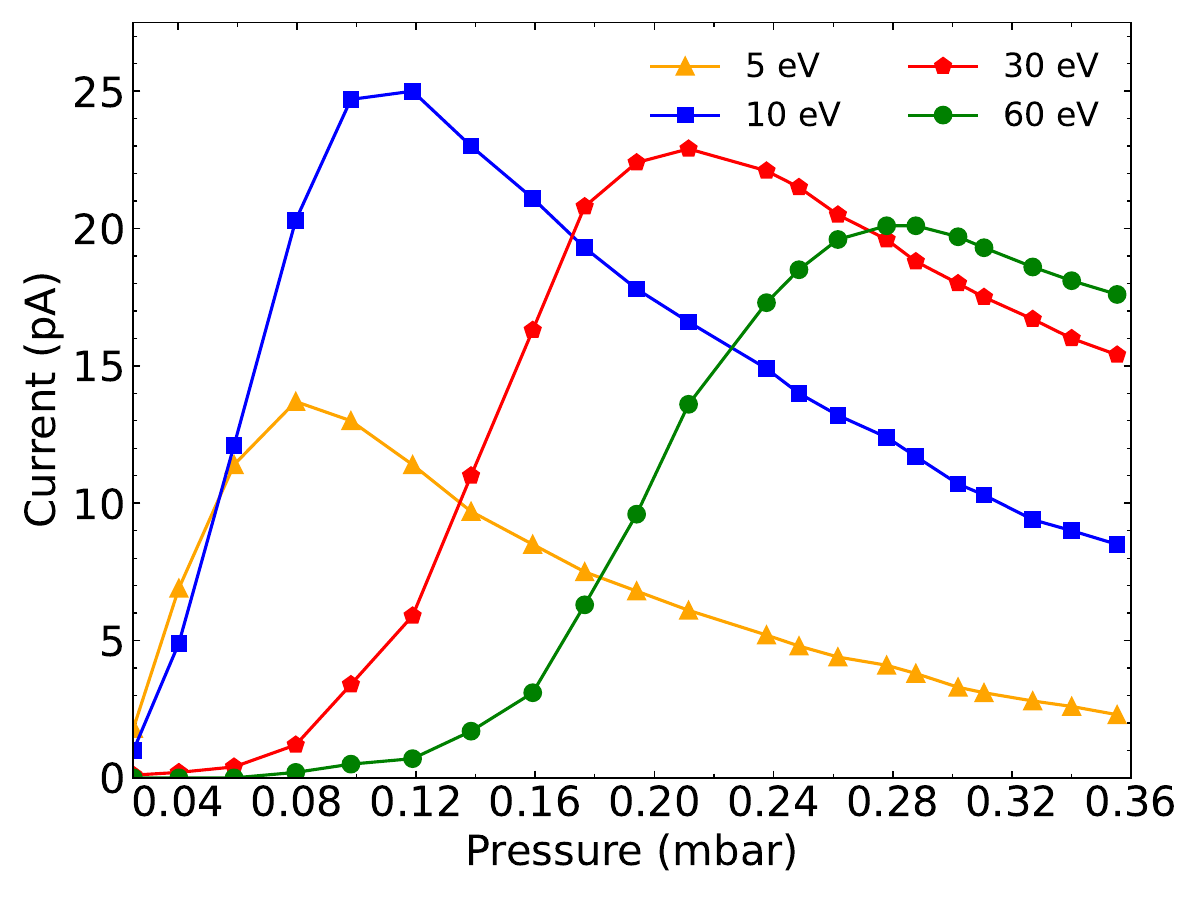}
\caption{Extraction current in continuous cooling mode as a function of buffer gas pressure inside the RFQ cooler cavity for beams with different injection energies ranging $\sim$5~eV to $\sim$60~eV, at a fixed $q$ parameter.\label{fig:pressure_injenergy}}
\end{figure}

With the $q$-parameter of the RF system fixed to maximize the transmission and the injection electrode voltages optimized to best decelerate the ions, the gas pressure in the RFQ cooler section was scanned between 0.02 to 0.36~mbar, to observe the changes in the transmitted ion current for different injection energies ranging from around 5~eV to 60~eV, see Fig.~\ref{fig:pressure_injenergy}. For a set injection energy, significant ion losses are to be accounted at too low pressure due to insufficient cooling. Their radial velocities remain too high to be efficiently transferred through the stage and ultimately through \textit{P1}. Increasing the pressure eventually leads to reaching a maximum extraction current, before dropping again. This drop at too high pressure can be explained by a shortening of the mean free path of the ions, ending up with a too small axial energy to be transferred towards the end of the RFQ cooler section. This decrease in current observed at high pressures can be countered with the use of a steeper gradient guiding the ions. However, the usage of a too high gradient also leads to ion losses. The optimum gradient value with which these tests were performed was found to be around 5~V/m on axis, see Tab.~\ref{tab:opvoltage}. 

The injection energy of the ions in the RFQ cooler section also has an influence on the optimum working gas pressure. The higher the injection energy, the more gas is needed to efficiently cool the charged particles and extract them. Buffer gas pressure of around 0.08~mbar is optimal for ions with $5$~eV injection energy where it is needed to be increased to around 0.30~mbar for 60~eV, see Fig.~\ref{fig:pressure_injenergy}. Moreover, the injection energy increase is not fully compensated by just increasing the buffer gas pressure, as evident in Fig.~\ref{fig:pressure_injenergy}. This is due to the smaller relative effect a change of pressure has at higher ion-atom collision rates. The generally lower transmitted currents at higher injection energies is due to higher working gas pressures, increasing the overall base pressure throughout the system as more gas gets outside of the RFQCB central chamber. Specifically on the injection side before the ions enter the RFQ cooler section, the ions are more likely to be lost via collisions with gas atoms outside of the RFQCB structure. In some degree increasing the pumping power can compensate for this effect. The best buffer gas cooling pressure for transmitting stable $^{85, 87}$Rb$^+$ ions was ultimately found to be around 0.12~mbar with ions injected into the RFQ cooler section with around 10~eV energy. Below this value, the ions do not have enough remaining kinetic energy to enter, thus the significantly lower transmitted currents observed at 5~eV. It is worth mentioning that these optimums are different for ions with a different mass-over-charge ratios.

\subsection{Longitudinal energy spread}
One of the main characteristics of the RFQ cooler stage is its ability to efficiently cool down the incoming ion distribution in order to reduce their energy both radially and axially, hence transferring a beam of better ion optical properties to the downstream experiments. The effect of the cooling process on the longitudinal energy spread $\Delta E_\text{long}$ of the continuous beam can be probed by scanning the voltage on the thin electrostatic plates \textit{injp} and \textit{P1} placed before the injection in the RFQ cooler and at its end, respectively, until the beam is fully blocked with high enough potential applied. The effective potential on the beam axis through the narrow apertures is calculated using electrostatic simulations performed with COMSOL Multiphysics \cite{multiphysics1998introduction}. 

\begin{figure}[htb!]
\includegraphics[width =\columnwidth]{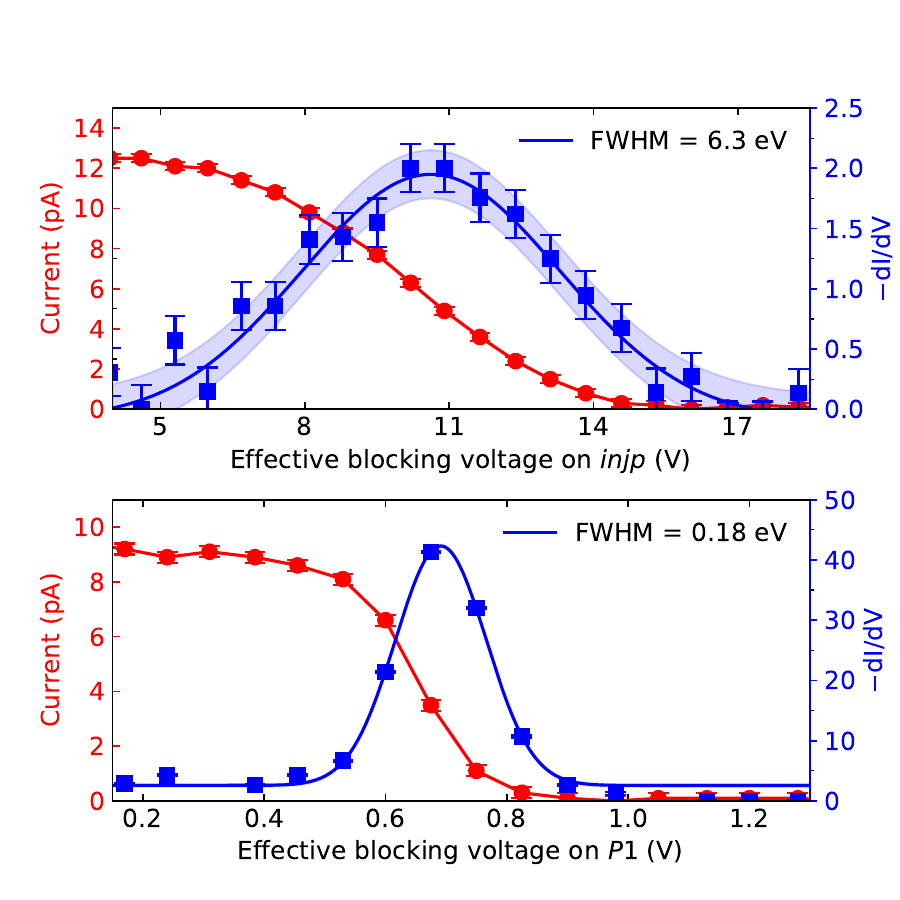}
\caption{Extraction current in continuous cooling mode, in red, for different effective blocking potential on-axis through \textit{injp} (top) and \textit{P1} (bottom). The red curves follow cumulative Gaussian distributions. The blue curves are obtained by differentiation of the current with respect to the applied voltage. The FWHM of the obtained Gaussians correspond to the longitudinal energy spread $\Delta E_\text{long}$ of the ions, entering the RFQ cooler through \textit{injp} (top), and exiting it through \textit{P1} (bottom). Incoming ions are decelerated to around 10~eV of kinetic energy upon injection in the RFQCB and cooled with $0.12$~mbar of buffer gas pressure. \label{fig:block}}
\end{figure}

The voltage applied on the plate is scanned around the $E/q$ of the beam while the transmitted ion current is recorded, see the obtained red curves in Fig.~\ref{fig:block}. They follow cumulative Gaussian distributions and $\Delta E_\text{long}$ is then obtained, in blue, from their differentiation with respect to the blocking voltage of \textit{injp} and \textit{P1}, plotted respectively on top and at the bottom of Fig.~\ref{fig:block}. The blue curves follow Gaussian distributions, which Full-Widths at Half-Maximum (FWHM) correspond to the $\Delta E_\text{long}$ of the ions passing through the diaphragm electrode. For stable $^{85, 87}$Rb$^+$ ions with 6~keV of kinetic energy decelerated down to around 10~eV when entering the RFQ cooler stage filled with 0.12~mbar of helium pressure, the $\Delta E_\text{long}$ is reduced from $\sim6$~eV upon injection down to $\sim0.2$~eV upon exit. The initial spread mainly originates from ion-atom collisions with the gas leaking to the injection beamline and increasing the energy dispersion before the ions are injected into the RFQCB through the \textit{injp} electrode. The $\Delta E_\text{long}$ upon extraction through \textit{P1} is representative of the spread of the cooled ions.

\subsection{Transmission efficiency}
\begin{figure}[htb!!]
\includegraphics[width =\columnwidth]{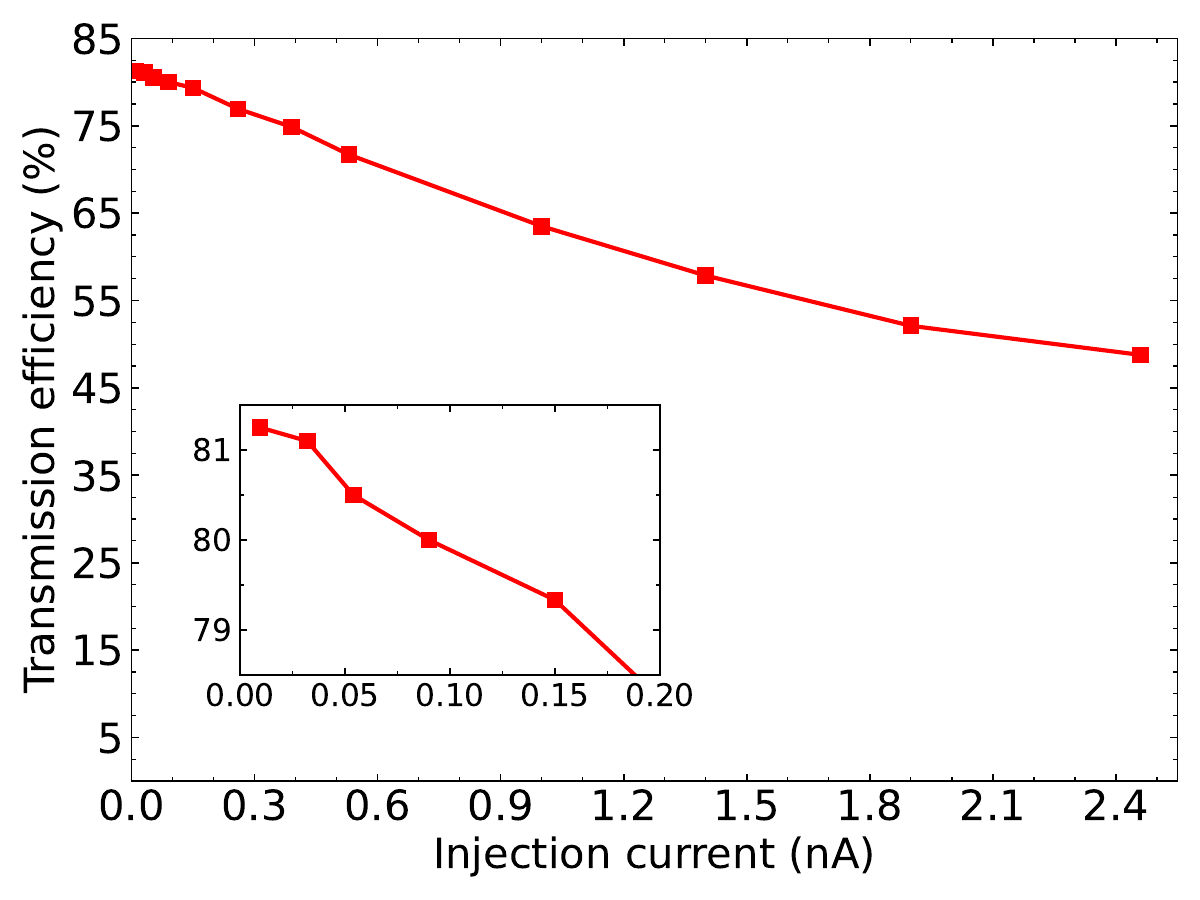}
\caption{Transmission efficiency in continuous transfer mode as a function of the injection current using the optimum cooling parameters aforementioned. The insert shows a close-up on the region of input current $<0.2$~nA.\label{fig:space_charge}}
\end{figure}

Using the optimum parameters determined and compiled here before, the transmission efficiency through the RFQCB in continuous cooling mode was determined for different injection beam intensities. The surface ionization source was heated gradually in order to produce from some pA to several nA of ion current. For each setpoint, the ion current was alternatively recorded on both FC placed in the injection and extraction cells, of same geometry and with the same suppressed voltage applied, by moving in and out of the beam axis the FC present on the injection actuator. The ion current injected in the RFQCB was varied from 0.01~nA to 2.45~nA. The corresponding transmitted currents are plotted in red in Fig.~\ref{fig:space_charge}.

At injection current intensities lower than $\sim0.15$~nA, the transmission in continuous cooling mode retains around 80$\%$ efficiency (see the insert in Fig.~\ref{fig:space_charge}). This result is in-line with existing RFQCB devices in other RIB facilities \cite{nieminen2001beam,mane2009ion,brunner2012titan,schwarz2016lebit,GERBAUX2023167631} and is suitable for the purpose of HIBISCUS, which is meant to deal with exotic beams and not high intensity currents. At injection currents $>0.15$~nA, the RFQCB starts to show saturation effects and have difficulties handling the increasing amount of charges that would need to be transferred through its structure. The higher heating currents applied on the ion source also contribute to this decrease of the transmission, as it increases the beam dispersion in space and energy and thus degrades the injection efficiency into the RFQ structure. However, these effects are gradual, and the RFQCB stills retain around 50$\%$ transmission efficiency for input currents of a few nA. 

\section{Performance in bunching mode}
This section reports on the first results of HIBISCUS in bunching mode. As a first and simpler demonstration of HIBISCUS' ability to release bunched beams, only the far-end of the RFQ cooler main section was used in this work, instead of the dedicated bunching sections and the sequence described in Sec.~\ref{modes}. This simple process was achieved by forming a potential well with the \textit{DC3} wedge electrodes (see Fig.~\ref{fig:cool}) and solely switching the \textit{P1} voltage between its "high" and "low" states to respectively accumulate the cooled ions and further release the them as bunches. In addition, the voltages in the RFQ buncher sections and the acceleration region were optimized to accommodate this different mode of operation, see Tab.~\ref{tab:opvoltagebuncher}.

\begin{table}[htb!]
 
    \centering
    \begin{tabular}{ll}
       Electrode & Voltage value (V) \\ \hline \hline
        \textit{RFQ cooler} &\\ \hline
        Injp & 6005\\
        DC1F/B & 5997.5/5993\\
        DC2F/B & 5996.5/5992\\
        DC3F/B & 5955/5910\\
        &\\
        \textit{RFQ bunchers} & \\ \hline
        1st segment & 5945\\
        2nd segment & 5950\\
        3rd segment & 5950\\
        4th segment & 5960\\
        5th segment & 5910\\
        6th segment & 5970\\
        &\\
        \textit{Switching plates} & \textit{Low}/\textit{High}\\ \hline
        P1 & 5695/6000\\
        P2 & 5925\\
        P3 & 5925\\
        P4 & 5890\\ 
        &\\
        \textit{Extraction optics} & \\ \hline
        Endcap & 5830\\
        Ext1 & 5000\\
        Ext2 & 4150\\
    \end{tabular}
    \caption{Typical HIBISCUS operating parameters in simple bunching mode using the far-end of the RFQ cooler stage for stable $^{85, 87}$Rb$^+$ ions with 6~keV of energy. Electrode labels are explicitly described throughout Sec.~\ref{exp}. The RF parameters, buffer gas pressure and injection voltages remain the same as in  Tab.~\ref{tab:oprf} and Tab.~\ref{tab:opvoltage}, respectively.}
    \label{tab:opvoltagebuncher}
\end{table}

\subsection{Capture of an ion sample in the RFQ cooler}
\begin{figure}[htb!!]
\includegraphics[width =\columnwidth]{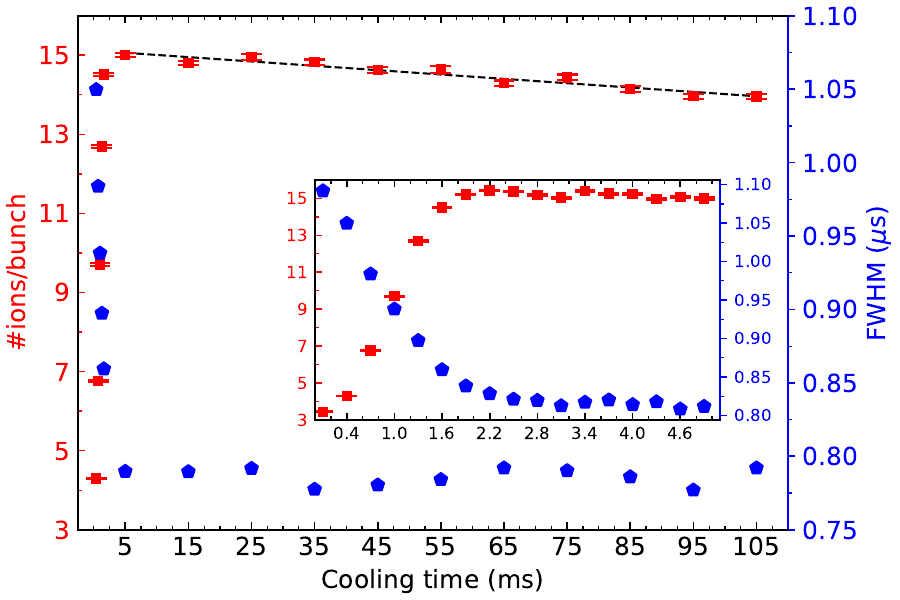}
\caption{Ion survival in the \textit{DC3} trapping region (in red) and FWHM of the extracted bunches (in blue) as a function of the cooling time $t_\text{cool}$ ranging from 0.1 to 105~ms (in red) in the \textit{DC3} trapping region using typical extraction parameters. The dashed gray linear trend corresponds to a rate loss of 0.07$\%$ per ms. The insert gives a closer look at the region between 0.1 and 5~ms.\label{fig:survival}}
\end{figure}

The ions stored in the trapping region at the far-end of the RFQ cooler (the \textit{DC3} region), continue to undergo elastic collisions with the buffer gas atoms, keeping them cooled at the bottom of the potential well. The optimum time needed before releasing the ion bunch can be determined, by varying the cooling duration and observing the temporal width of the extracted bunch. This was performed by first opening the beamgate for 2~ms, to inject some ions into the RFQCB. They were then let to cool down and further captured in the \textit{DC3} region with \textit{P1} in its high-state for a set time $t_\text{cool}$. \textit{P1} was finally switched to its low-state for 100~$\mu$s to allow the bunch to be extracted. After acceleration, the temporal structure of the ion bunches was ultimately recorded with the MagneToF in the extraction cell. The detector efficiency was not taken into account for this study. The blue curve in Fig.~\ref{fig:survival} was obtained, for $t_\text{cool}$ ranging from 0.1 to 105~ms. It shows the minimum duration needed to have the ion bunch efficiently cooled at the bottom of the potential well before its extraction. Thus, for $t_\text{cool}\geqslant2.5$~ms, roughly constant bunch width around FWHM $=0.8$~$\mu$s is observed, while shorter times lead to a wider temporal distribution..

In this work, the survival of the ion bunch in the \textit{DC3} trapping region was also studied, as the use of longer capture times can lead to significant ion losses. These can be attributed to neutralization processes, molecular formation with impurities present in the buffer gas, or motion instability causing the particle to hit the electrodes structure. In the present case, with stable $^{85, 87}$Rb$^+$ ions, low charge densities captured and the RFQCB structure filled with ultra-high-purity helium, the main factor inducing losses is expected to be the instability of the ion confinement, which can thus be characterized. On the red curve shown in Fig.~\ref{fig:survival}, it is observed that the extracted ion rate reaches its maximum at 2.5 ms of cooling. Beyond this the rate slowly decreases, following a linear loss trend of 0.07$\%$ per ms. With this process, the bunching efficiency using optimum timings and parameters, and considering some tens of ions per bunch lies around 80$-$90$\%$ compared to HIBISCUS operating in continuous cooling mode.

\subsection{Bunch extraction from the RFQ cooler}
\begin{figure}[htb!]
\includegraphics[width =\columnwidth]{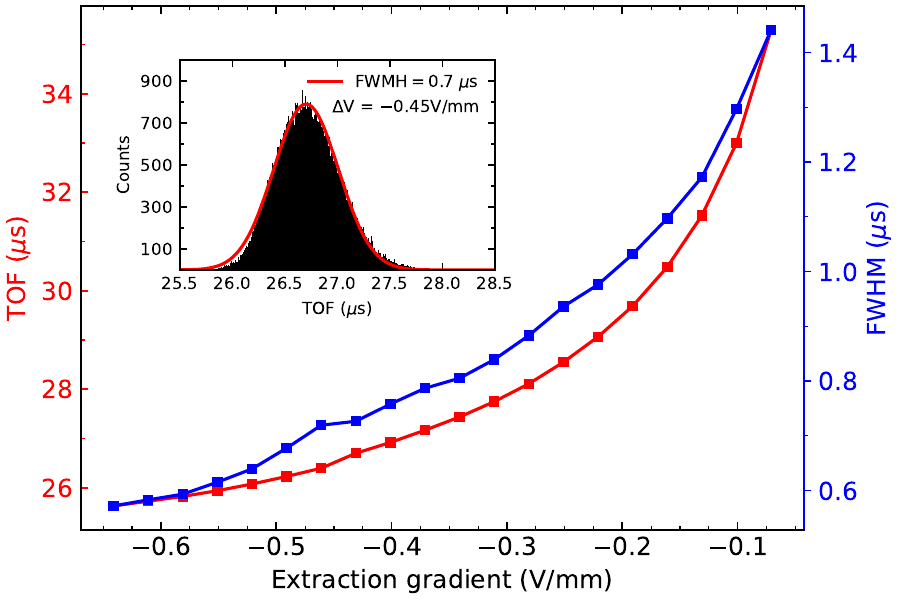}
\caption{Effect of the extraction gradient in V/mm on the FWHM of the extracted bunches (in blue) and the mean ion-TOF between their release and detection (in red), in $\mu$s. The ion sample was kept in the RFQ cooler section for 2.5~ms before extraction. The insert gives an example of the bunch Gaussian time structure for an extraction gradient of -0.45~V/mm.\label{fig:extractionP1}}
\end{figure}

In addition to its preparation, the ion bunch extraction also needs to be tuned according to the beam requirements of the experimental setups placed downstream. Depending on the voltage applied when switching \textit{P1} between its "high" and "low" states, the ions will gain more or less velocity upon extraction from the trapping region. These so-called fast or slow extractions respectively favor a reduced temporal width or a minimized longitudinal energy spread of the ion sample recorded on the detector. The effect of the extraction voltage on the time spread was especially looked at in this work, by varying the "low" state voltage applied on \textit{P1} from 65 to 465~V, giving a gradient $\Delta V$ ranging from $-$0.07 to $-$0.67~V/mm. The corresponding mean ion-TOF of arrival on the detector and the FWHM of the ion bunch are shown in Fig.~\ref{fig:extractionP1} with the red and blue curves, respectively. Using small gradients, the ion sample going through \textit{P1} has more time to diverge in time, giving a wider temporal distribution and a slower drift towards the detector. With increasingly steeper gradient on the contrary, the bunch is extracted faster and gets narrower in time, reaching a FWHM as low as 0.6~$\mu$s for $\Delta V\sim-0.6$~V/mm. A Gaussian time structure of a typical bunch is shown in the insert in Fig.~\ref{fig:extractionP1}, for $\Delta V=-0.45$~V/mm. Finally, no significant difference in efficiency was observed between the application of a fast or a slow extraction, making HIBISCUS equally adequate for the use of both.

\section{Summary and outlook}
A new RFQCB device named HIBISCUS has been developed and commissioned offline in view of its implementation at the MATS-LaSpec facility and as an in-kind contribution from Finland for the FAIR project. The characteristics of the RF and the buffer gas feeding systems, together with the performance of the experimental setup have been studied and optimized using 6~keV beams of stable $^{85, 87}$Rb$^+$ ions. Using the continuous cooling mode, the ions extracted from the RFQ cooler stage of HIBISCUS were shown to have their longitudinal energy spread reduced down to $<1$~eV with an 80$\%$ efficiency for current intensities up to $150$~pA  and 50$\%$ up to 2~nA. The far-end of the RFQ cooler section can as well be used to capture and release ions by delaying their extraction using electrodes with fast-switching voltages. Using this process, HIBISCUS is able to produce ion samples with sub-microsecond width while remaining transport efficient. The stability and efficiency of the ion confinement was also demonstrated. These results make HIBISCUS suitable for preparing ion beams for high precision mass spectrometry or post-trap decay spectroscopy with Penning traps, and collinear laser spectroscopy experiments.

More generally, further developments are on their way to commission the RFQ buncher sections of HIBISCUS to also fit with experimental setups having more demanding beam requirements, such as multi-reflection time-of-flight mass spectrometers (MR-TOF-MS). Especially, these trapping regions will allow reduce the time spread of the extracted bunches at the time focus, reaching temporal width down to $\sim100$~ns. Furthermore, a diagnostic tool to study the longitudinal energy spread of the extracted bunches will also be installed in the extraction cell in the future (see Fig.~\ref{fig:diagnostics}). This will be of use to characterize how the extraction parameters affect the spread, allowing it to be optimized according to the requirements of collinear laser spectroscopy. 

Finally, given the diversity of exotic beams that will be available at the MATS-LaSpec facility, more extensive studies of the performances of HIBISCUS will be performed offline in different mass ranges of operation to ensure its versatility.

\section*{Acknowledgments}
We acknowledge the financial support provided by Helsinki Institute of Physics, and the European Union’s Horizon 2020 research and innovation programme under Grant Agreement No. 771036 (ERC CoG MAIDEN). The Vilho, Yrjö and Kalle Väisälä Foundation is acknowledged by J. Ruotsalainen. T. Eronen acknowledges support from the Research Council of Finland under grant number 295207, 306980 and 327629. A. Kankainen acknowledges support from the Research Council of Finland under grant number 354968.

\appendix





\bibliographystyle{elsarticle-num} 
\bibliography{bibtex}






\end{document}